\documentclass[10pt]{article}
\usepackage{sao2}
\usepackage{psfig,epsf}

\issue{2011, 66, 171--182}

\begin{document}
\markboth{
O.V.~Verkhodanov, M.~Singh, A.~Pyria, S.~Nandi,
N.V.~Verkhodanova}
{A STUDY OF GIANT RADIO GALAXIES AT RATAN-600}
\title{A Study of Giant Radio Galaxies at RATAN-600}

\author{M.L.~Khabibullina\inst{a},
O.V.~Verkhodanov\inst{a},
M.~Singh\inst{b},
A.~Pirya\inst{b},
S.~Nandi\inst{b},
N.V.~Verkhodanova\inst{a}
}
\institute{
$^a$\saoname; \\ $^b$\ARIES}

\date{July 28, 2010}{September 15, 2010}
\maketitle
\begin{abstract}
We report the results of flux density measurements in the extended
components of thirteen giant radio galaxies, made with the
RATAN-600 in the centimeter range. Supplementing them with the
WENSS, NVSS and GB6 survey data we constructed the spectra of the
studied galaxy components. We computed the spectral indices in the
studied frequency range and demonstrate the need for a detailed
account of the integral contribution of such objects into the
background radiation.

\end{abstract}
\maketitle

\section{INTRODUCTION}

According to the generally accepted definition,	  giant radio
galaxies (GRGs) are the radio sources with linear sizes greater
than 1\,Mpc, i.e. the largest radio sources in the Universe. They
mostly belong to the morphological type FR\,II
\cite{Fanaroff} and are identified with giant
elliptical galaxies and quasars. In comparison with normal
galaxies, GRGs are quite rare. This makes their statistical study
and a detailed research into the causes of their formation as a
population quite difficult. They are the biggest objects in the
visible universe, and it is possible that they could play a
special role in the formation of the large-scale structure. Radio
observations of GRGs allow to clarify what caused such gigantic
objects to appear. Large dimensions of GRGs as well suggest that
these sources must be at the last stages of evolution.

\begin{figure*}[!tbp]
\centerline{ \vbox{
\hbox{
 \psfig{figure=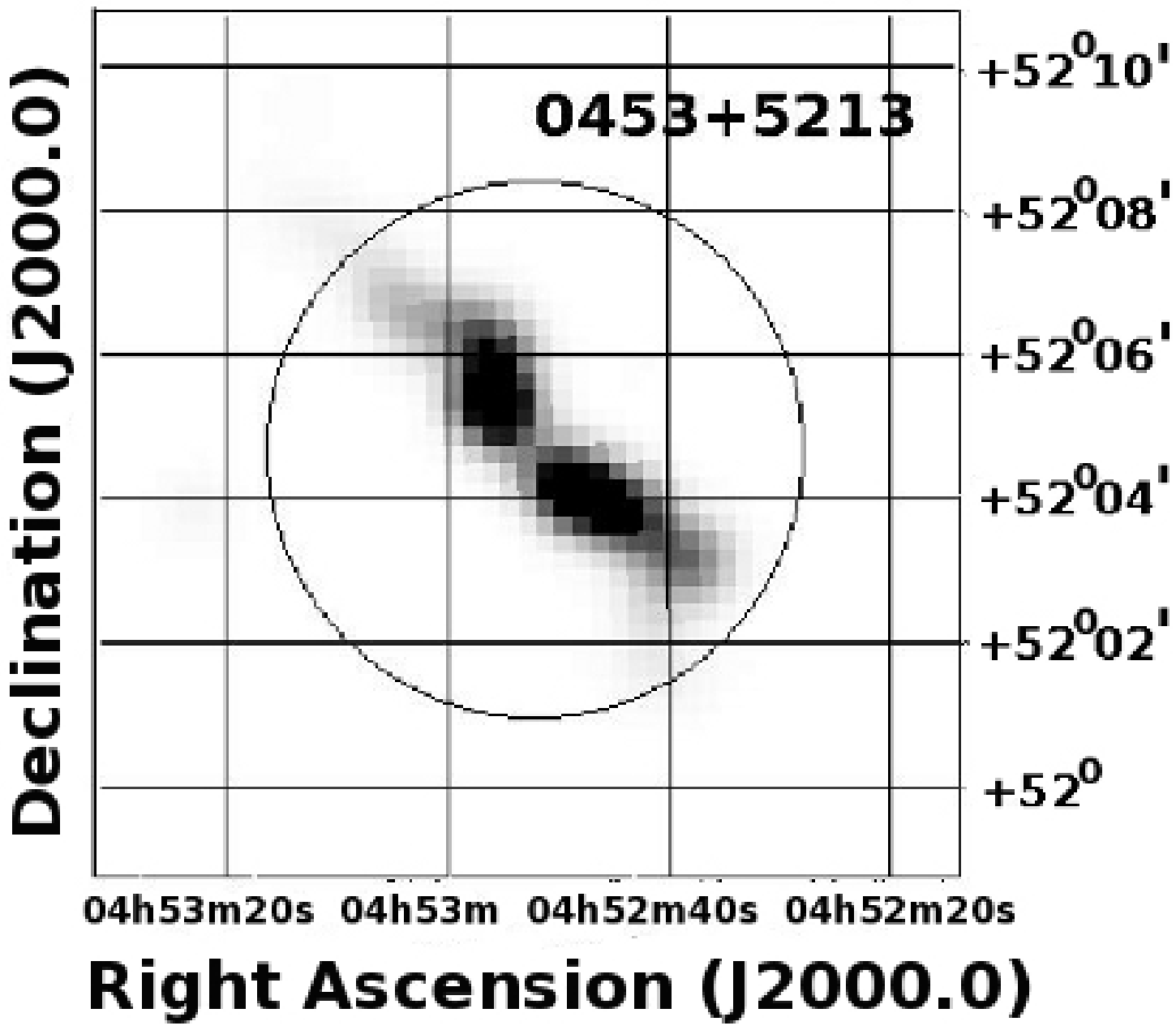,width=6cm,angle=-90}
 \psfig{figure=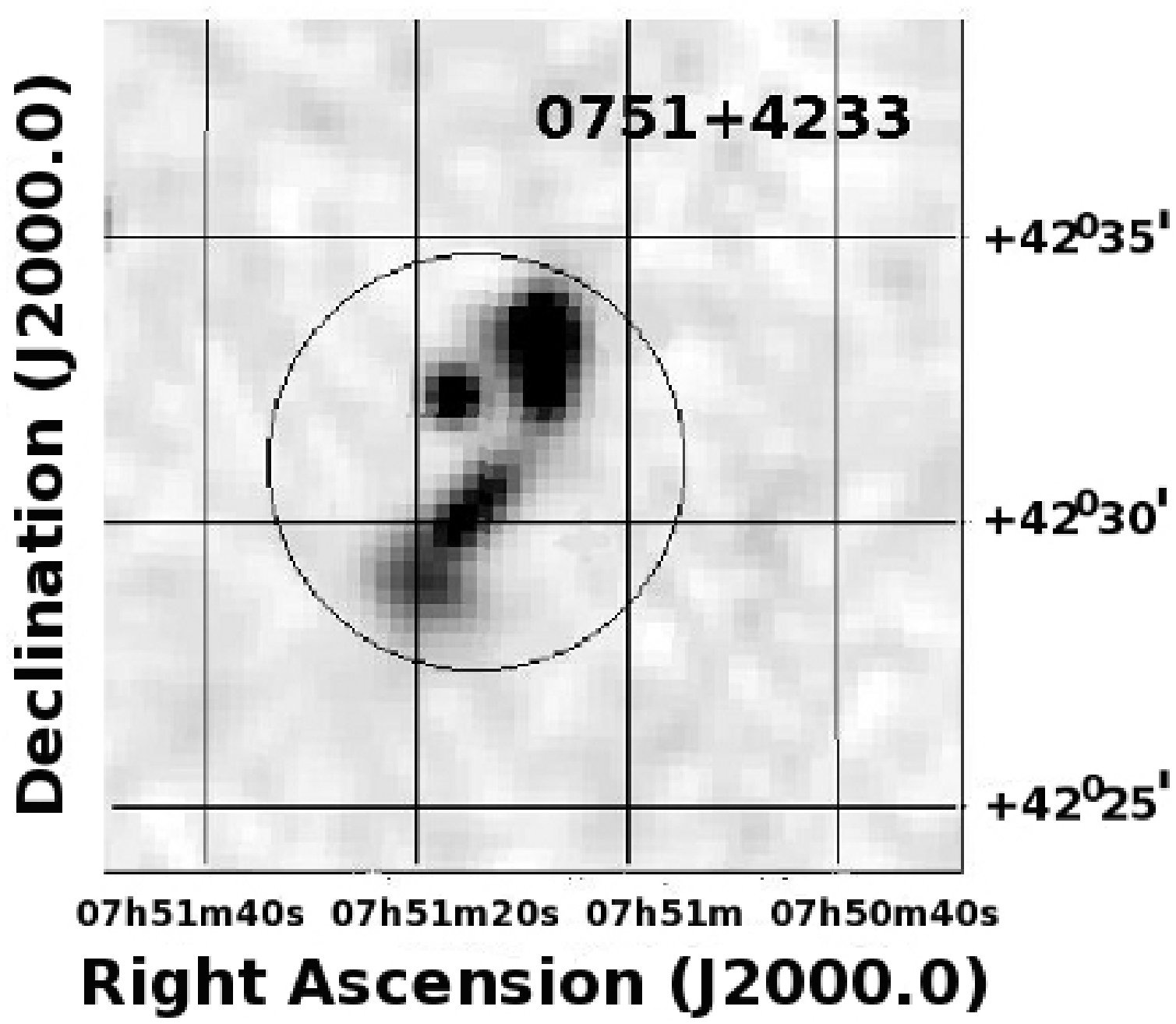,width=6cm,angle=-90}
} \hbox{
\psfig{figure=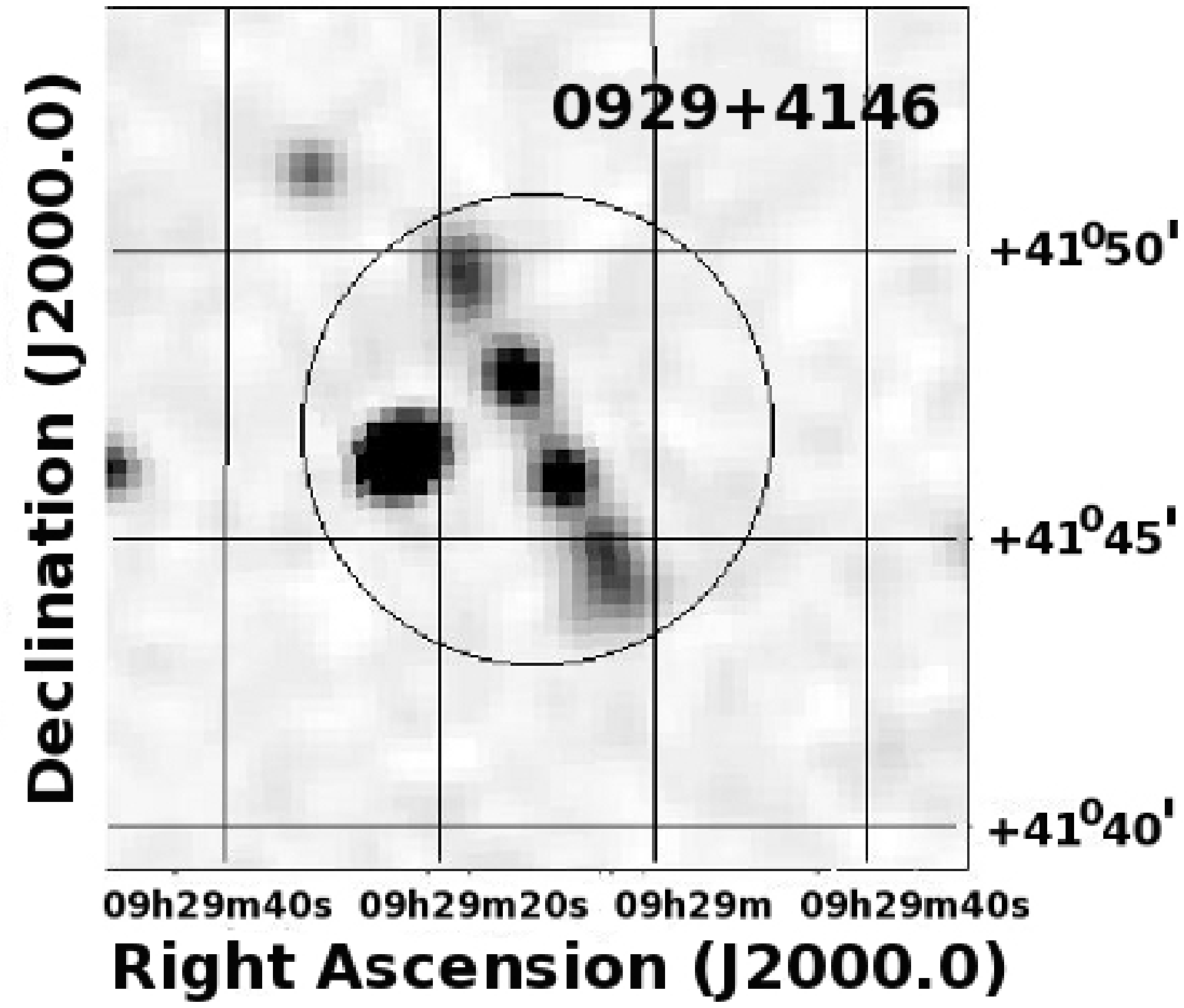,width=6cm,angle=-90}
\psfig{figure=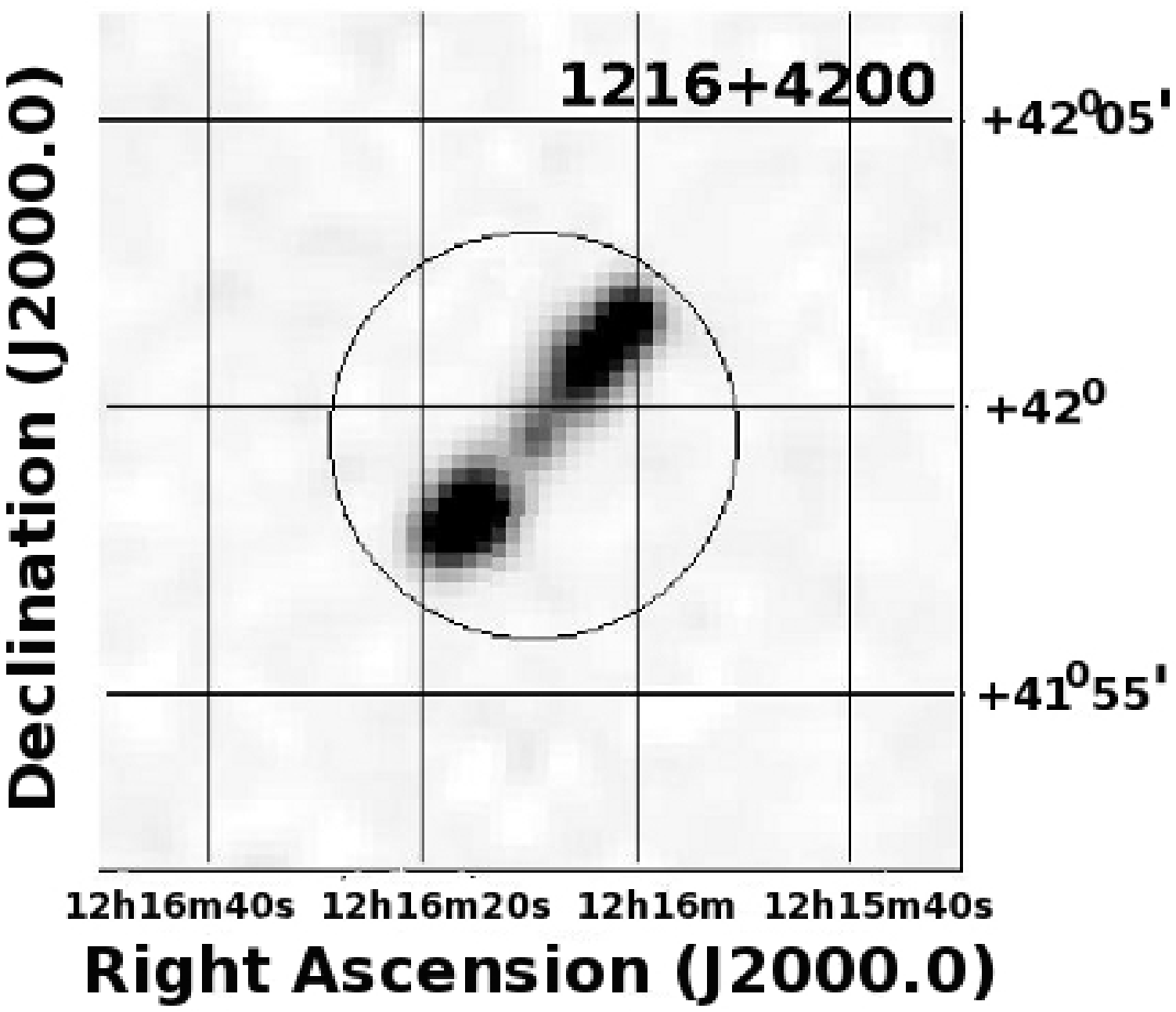,width=6cm,angle=-90}
} \hbox{
\psfig{figure=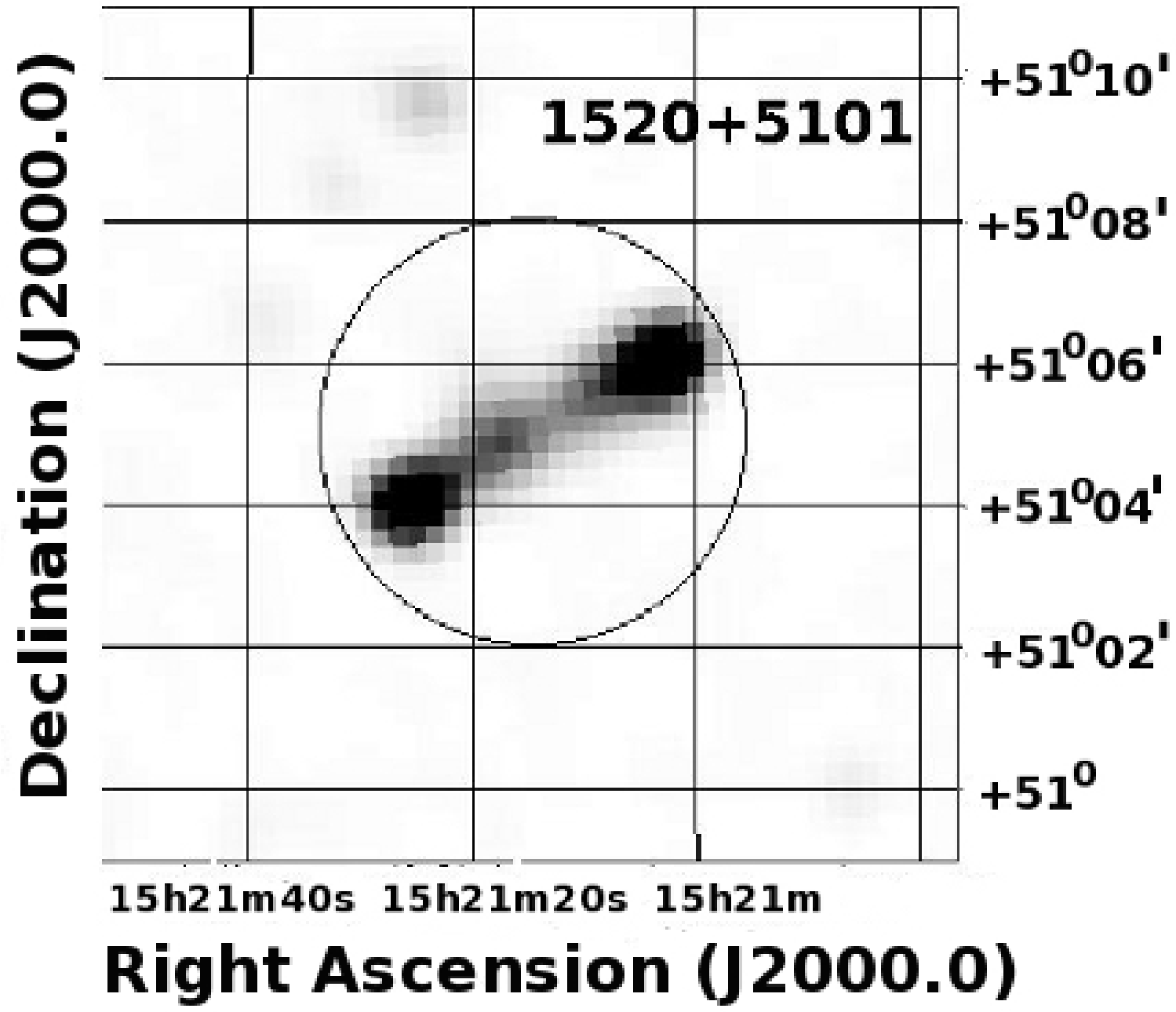,width=6cm,angle=-90}
\psfig{figure=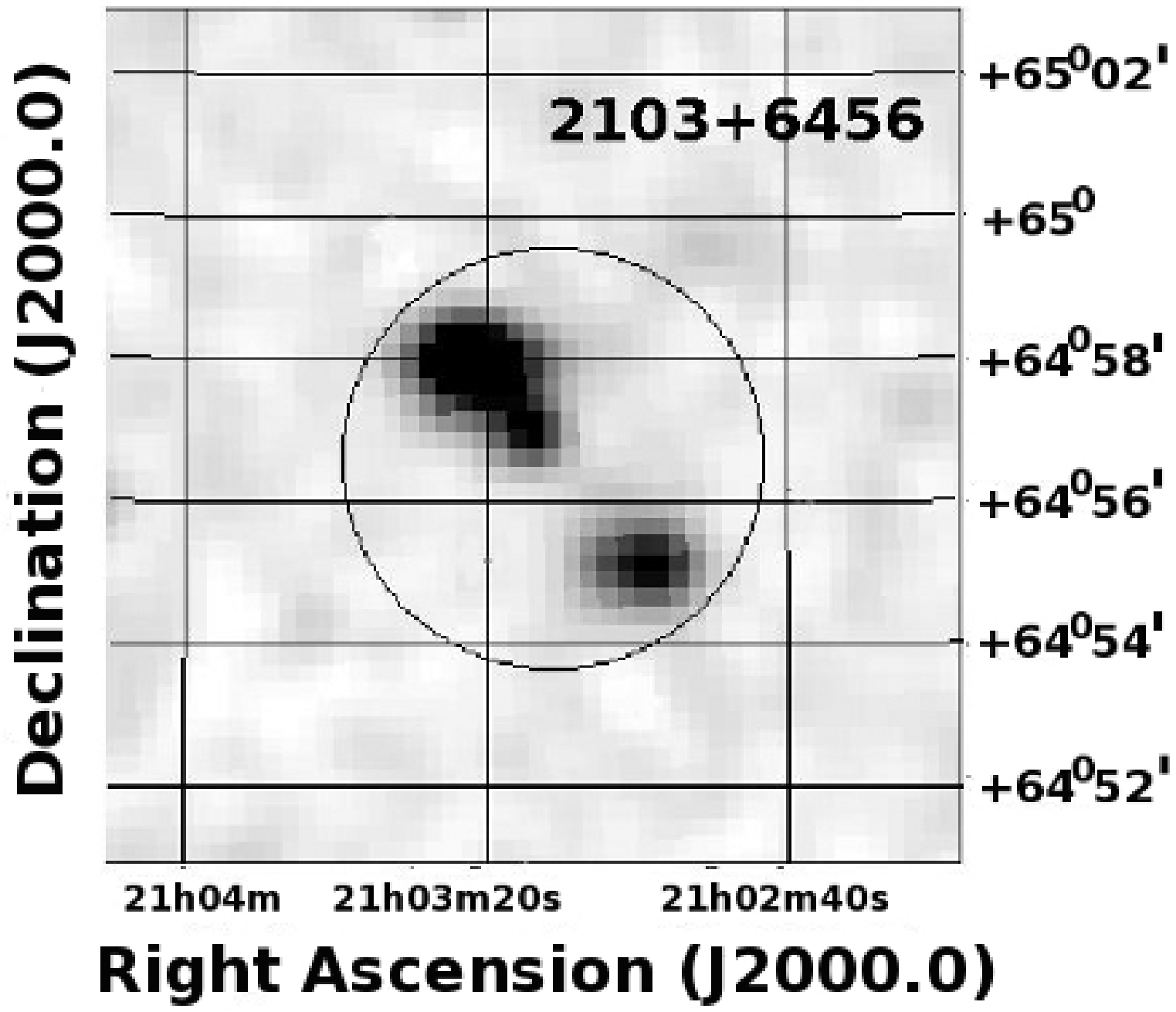,width=6cm,angle=-90}
} \hbox{
\psfig{figure=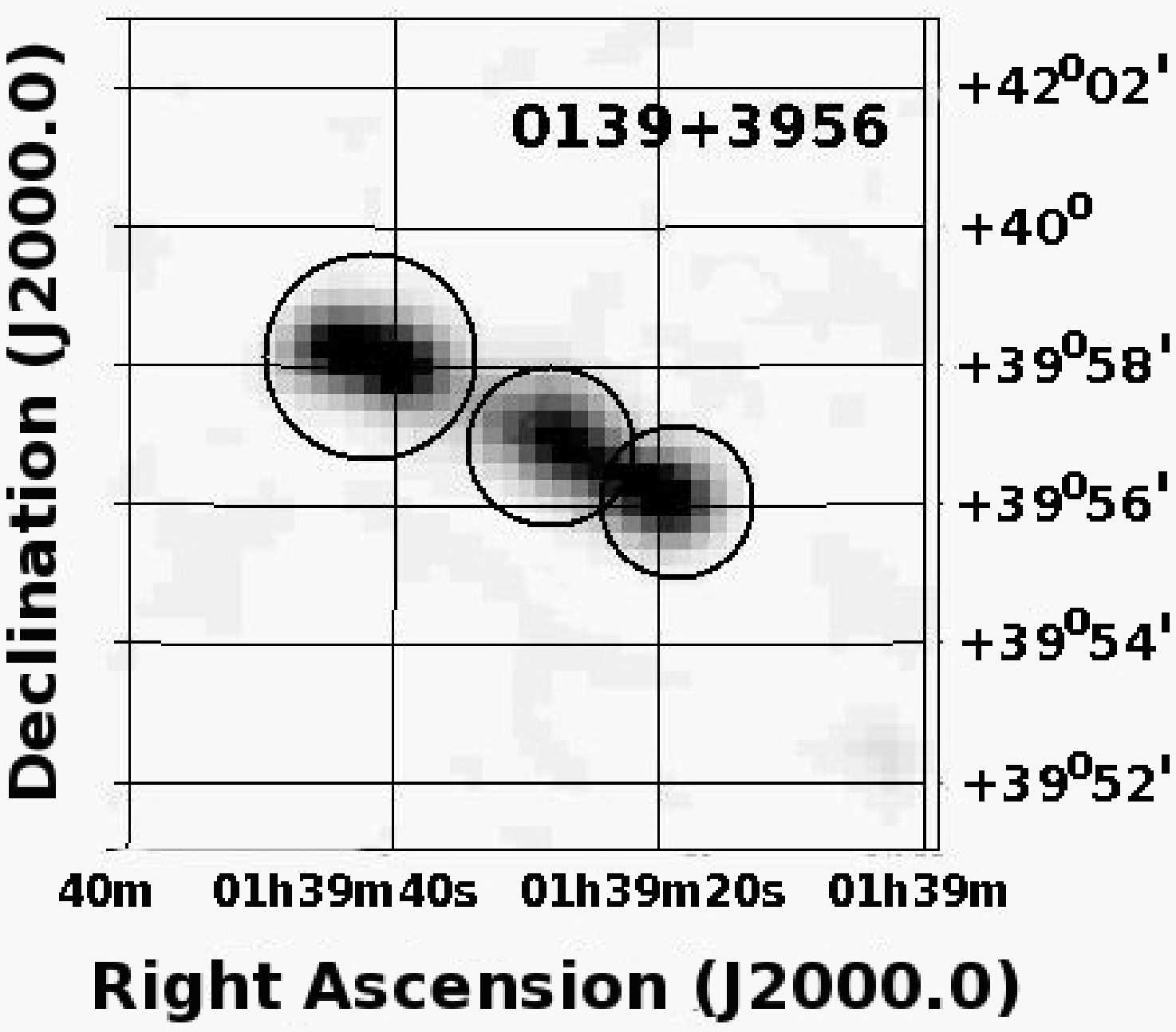,width=6cm,angle=-90}
\psfig{figure=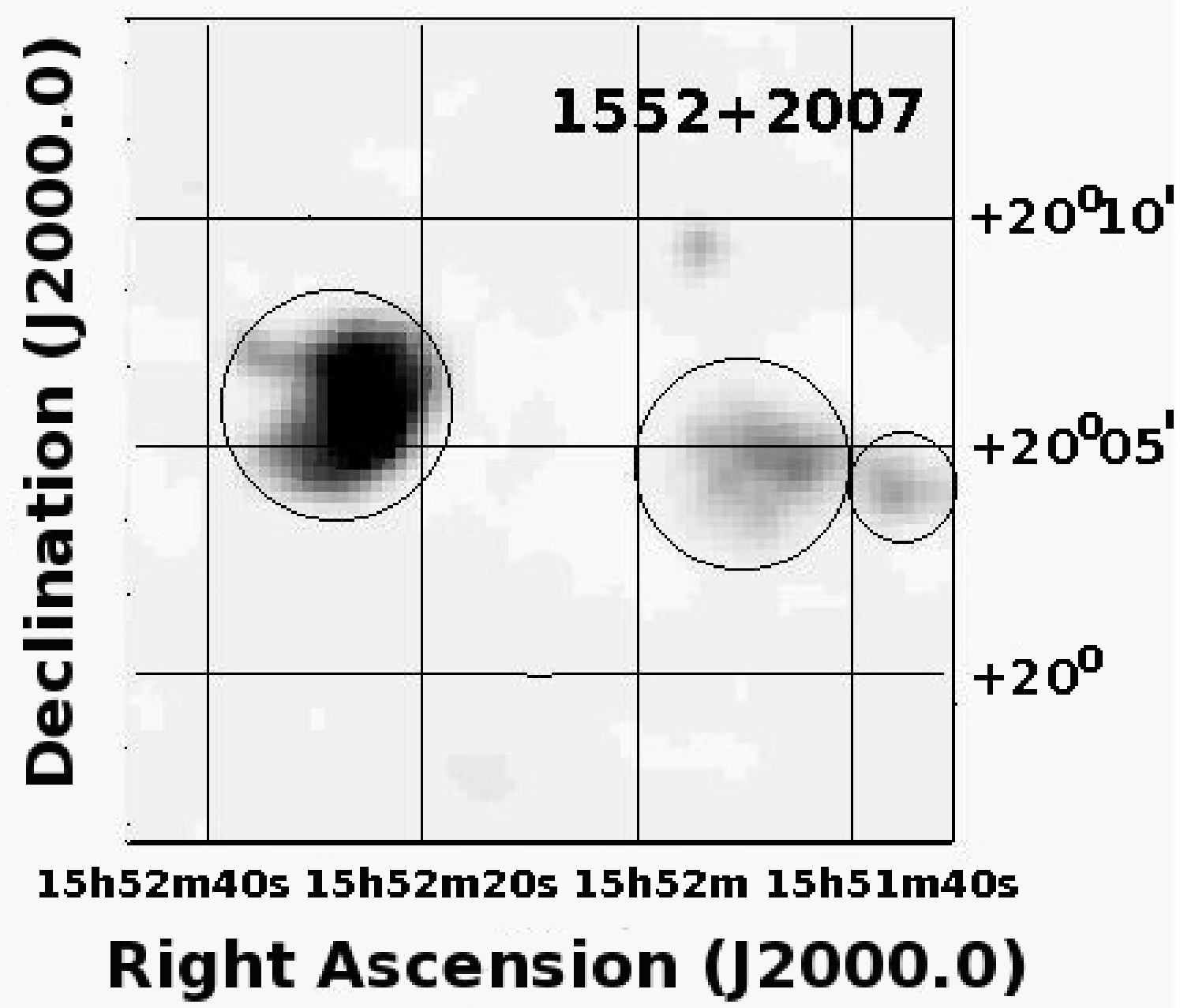,width=6cm,angle=-90}
} } } \caption{Radio images of giant radio galaxies in the NVSS
survey. The circles mark the object components, observed with the
RATAN-600.} \label{ff1a:Verkhodanov_n}
\end{figure*}

\begin{figure*}[!tbp]
 \centerline{ \vbox{
 \hbox{
  \psfig{figure=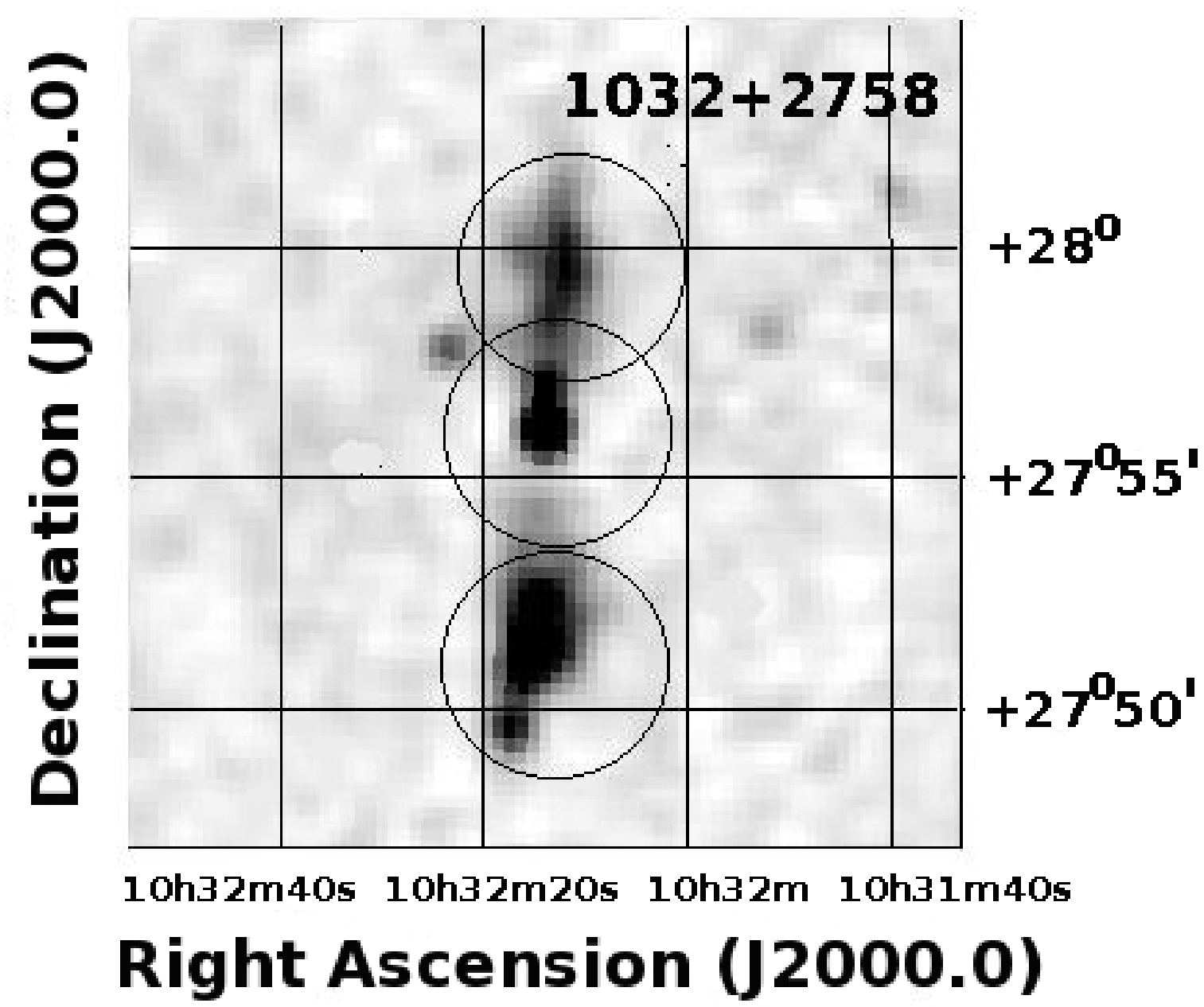,width=6cm,angle=-90}
  \psfig{figure=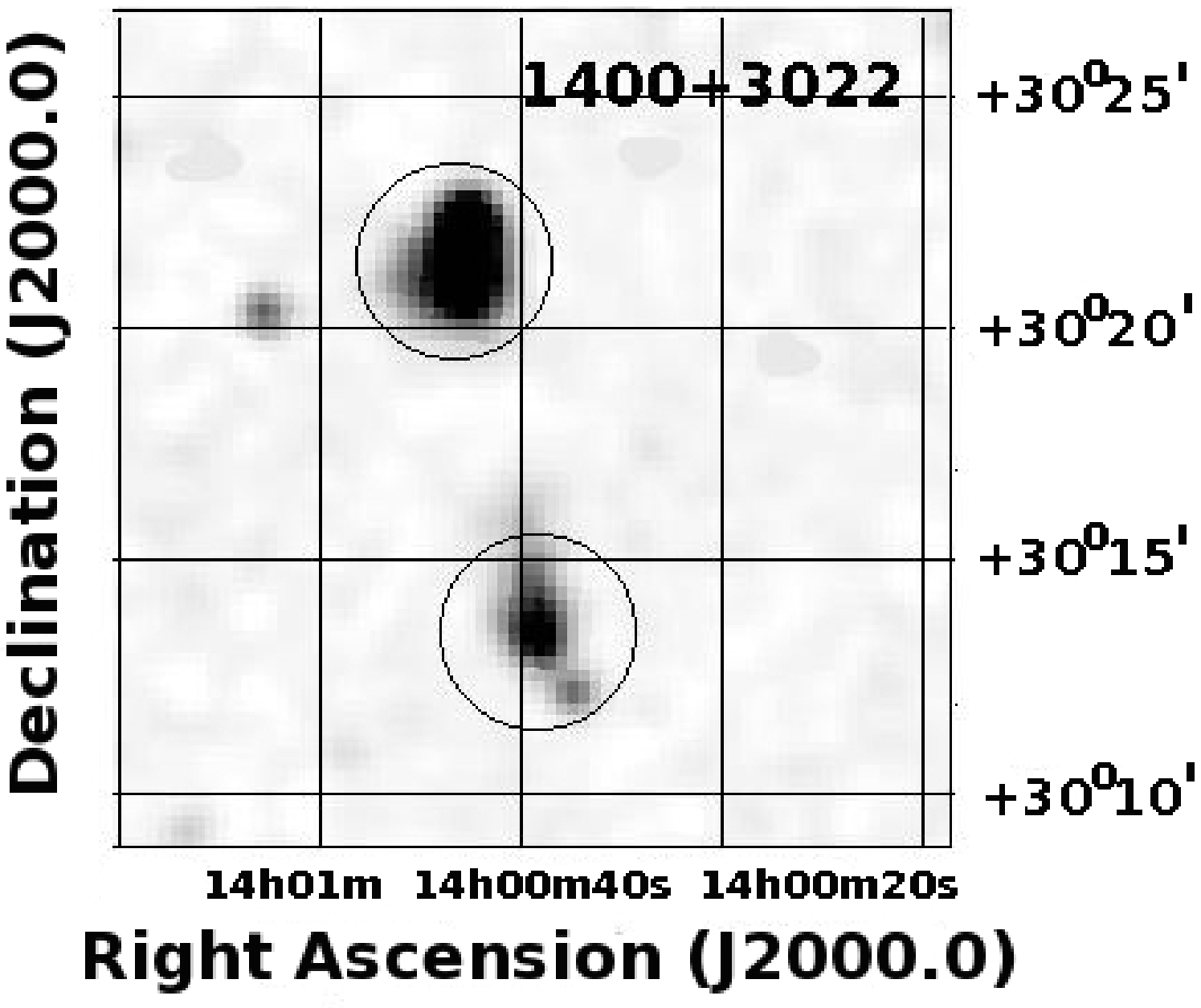,width=6cm,angle=-90}
} \hbox{
 \psfig{figure=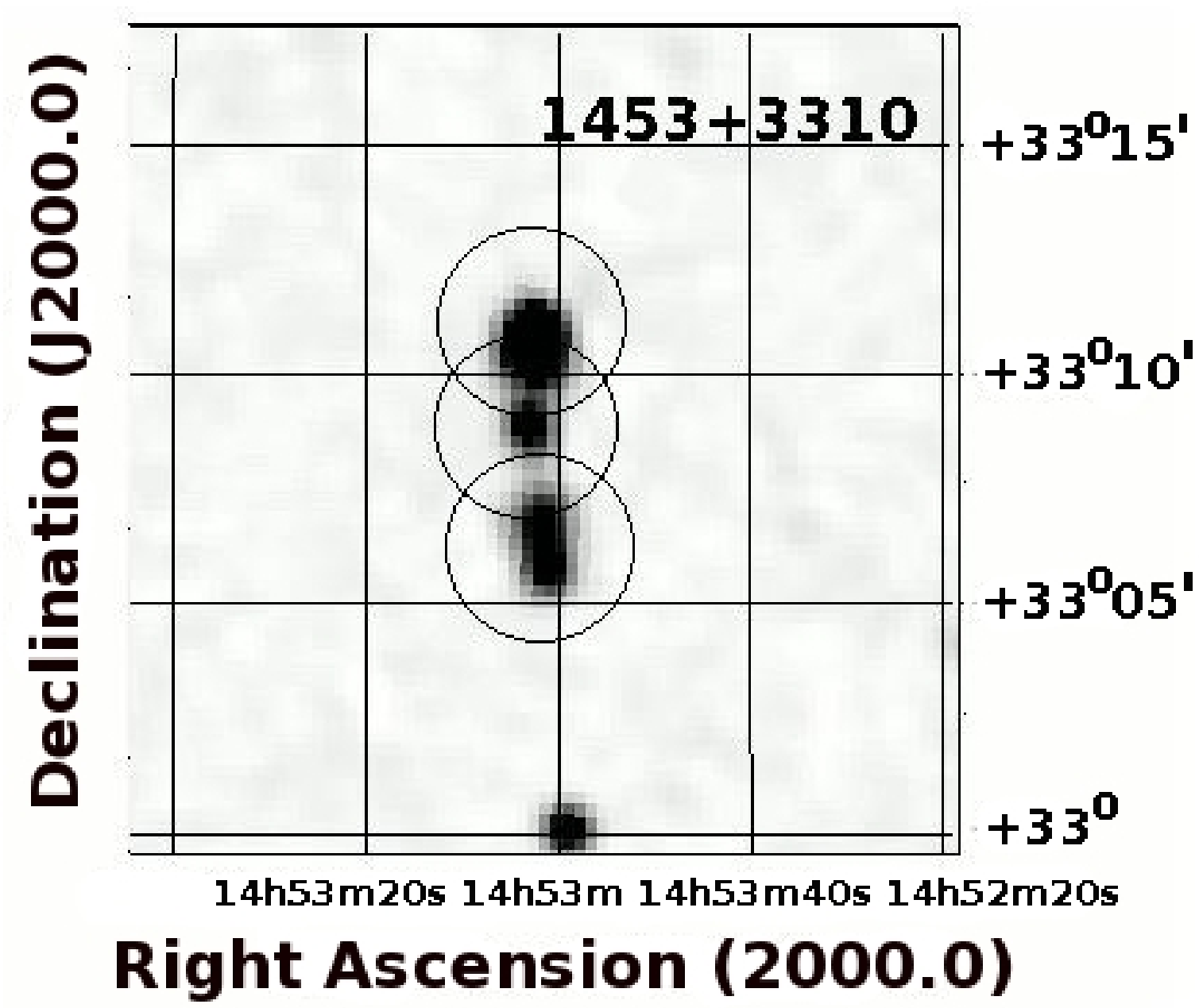,width=6cm,angle=-90}
 \psfig{figure=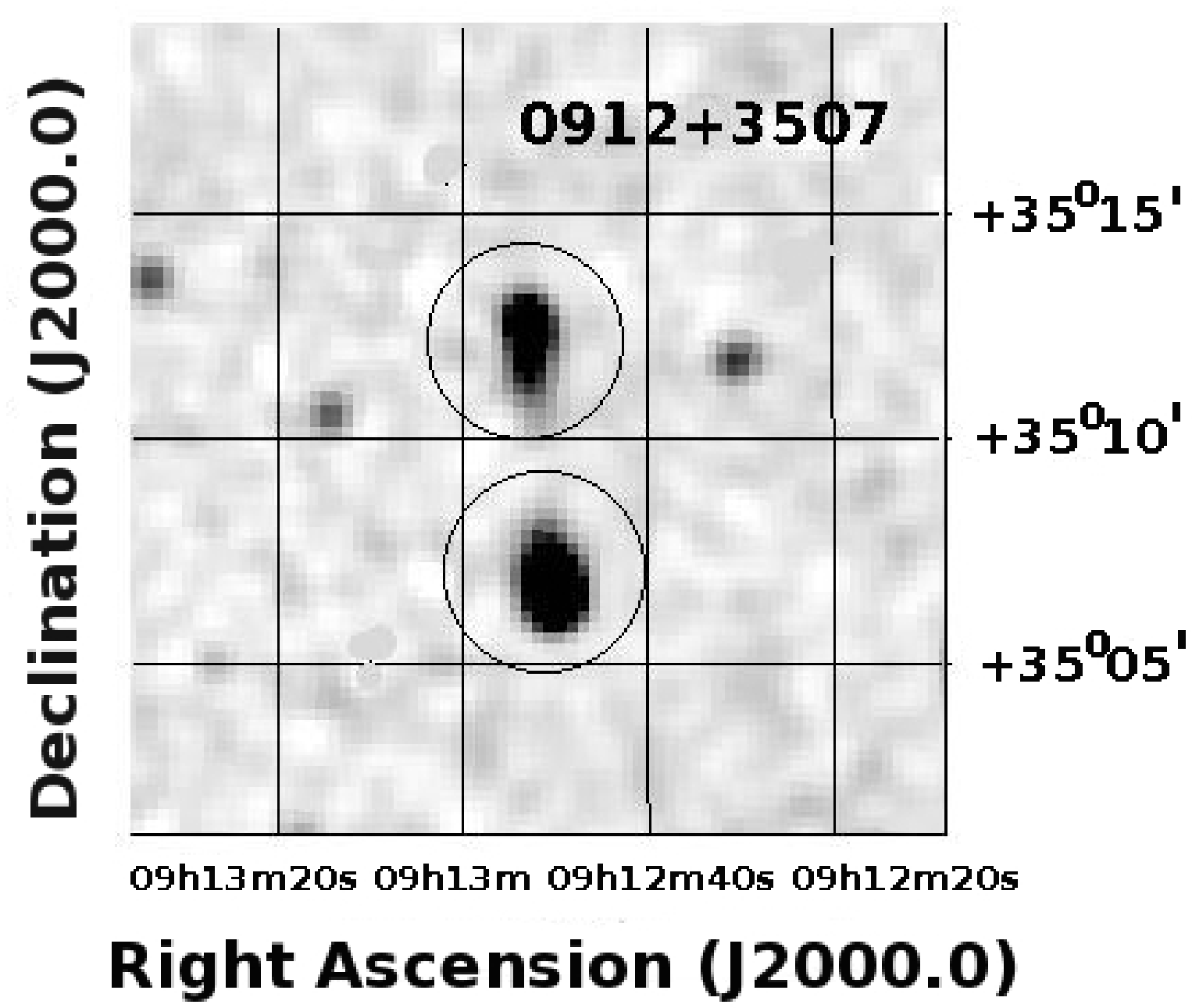,width=6cm,angle=-90}
} \hbox{
 \psfig{figure=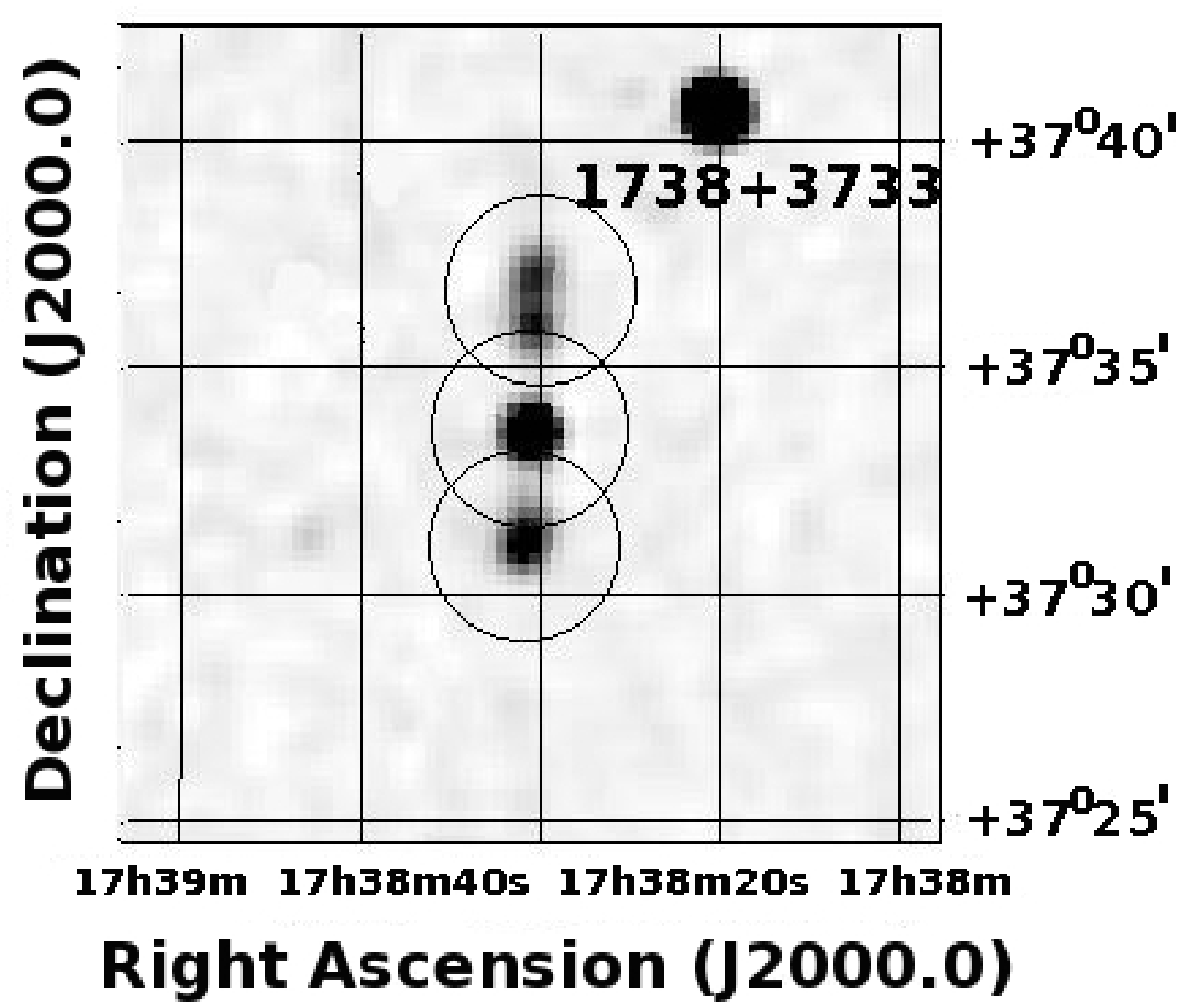,width=6cm,angle=-90}
  \mbox{\hspace*{8cm}}
} }}
{ {\bf Fig.\,1.} (Contd.) }
\end{figure*}

The study of these objects began with the source 3C\,236
\cite{Strom}. The models of radio sources
\cite{Kaiser_Dennett},\cite{Blundell}
predict temporal variations in the radio luminosity and linear
sizes of giant radio sources. According to these models, GRGs
should be very old objects (aged over $10^8$ yrs) that are
presumably located in the environments with decreased matter
density, as compared to the smaller sources with comparable
radio luminosities \cite{Kaiser_Alexander}.
Komberg and Paschenko \cite{Komberg} have analyzed
the radio and optical data (SDSS, APM) for the radio galaxies and
quasars, and concluded that apart from the effect of the
environment, the giant sizes of radio sources can be attributed to
the population of long-lived radio-loud active nuclei, which in
turn can evolve into to GRGs. Multi-frequency observations
\cite{Mack} have shown that the spectral age of GRGs
is longer than the one, expected from the evolutional models. As
noted in \cite{Jamrozy_Machalski}, such radio
galaxies may ~affect the processes of galaxy formation, since the
pressure of gas, outflowing from the radio source, may compress
the cold gas clouds thus initiating the development of stars on
the one hand, and stop the formation of galaxies on the other
hand. Several groups
\cite{Schoenmakers_Mack}-\cite{Nandi}
continue to study the characteristics of GRGs, longing to explain
their gigantic sizes. However,	no unambiguous solution of this
problem exists up to date.

\begin{table*}[!tbp]
\begin{center}
 \caption{Main
parameters of the observed giant radio galaxies}
\begin{tabular}{c|c|c|c|r|r}
\hline
 Source		 &   Coordinates    & Redshift & Type & Angular & Flux density \\
     & RA+Dec (J2000.0)& &     & size, &   (1.4\,GHz),	\\
	  & hhmmss+ddmmss   &	      &	    & arcmin& mJy  \\
\hline
GRG 0139+3957 & 013930+395703	&  0.211  & II	&  5.7	  &  801.1   \\
GRG 0452+5204 & 045253+520447	& 0.109	  & I	& 9.7	  & 2869.1  \\
GRG 0751+4231 & 075109+423124	& 0.203	  & II	& 6.0	  & 162.3   \\
GRG 0912+3510 & 091252+351016	&  0.249  & II	&  6.2	  &  157.4   \\
GRG 0929+4146 & 092911+414646	& 0.365	  & II	& 6.6	  & 165.5   \\
GRG 1032+2759 & 103214+275600	&  0.085  & II	&  11.0	  &  284.1   \\
GRG 1216+4159 & 121610+415927	& 0.243	  & II	& 5.2	  & 415.5   \\
GRG 1343+3758 & 134255+375819	&  0.227  & II	&  11.3	  &  131.0   \\
GRG 1400+3017 & 140040+301700	&  0.206  & II	&  10.8	  &  451.9   \\
GRG 1453+3308 & 145303+330841	&  0.249  & II	&  5.7	  &  455.5   \\
GRG 1521+5105 & 152115+510501	& (0.37)  & II	& 4.3	  & 1197.5  \\
GRG 1552+2005 & 155209+200524	&  0.089  & II	&  19.6	  &  2385.6  \\
GRG 1738+3733 & 173821+373333	&  0.156  & II	&  6.5	  &  236.0   \\
GRG 1918+516~~~& 191923+514334	 & 0.284   & II	 & 7.3	   & 920     \\
GRG 2103+6456 & 210314+645655	& 0.215	  & II	& 4.8	  & 119.7   \\
\hline
\end{tabular}
\end{center}
\end{table*}

\begin{table*}[!tbp]
\begin{center}
\caption{The observed
GRG regions. Sections: c---central, n---northern, s---southern.
$N_t$ is the number of transits. The coordinates (right
ascension+declination) of the centers of components are listed for
the epoch J2000.0}
\begin{tabular}{c|c|c|r}
\hline
Source		  &Section&   Coordinates of center& $N_t$ \\
	   &	   &	observed regions   &	   \\
\hline
GRG 0139+3957  & c     &    013927.4+395653 &  1    \\
\hline
GRG 0452+5204  & c     &    045343.7+520556 &  11   \\
\hline
GRG 0751+4231  & c     &    075153.9+422945 &  10   \\
\hline
GRG 0912+3510  & n     &    091252.0+351231 &  5    \\
GRG 0912+3510  & s     &    091250.0+350631 &  1    \\
\hline
GRG 0929+4146  & c     &    092951.8+414353 &  10   \\
\hline
GRG 1032+2756  & n     &    103212.5+275925 &  3    \\
GRG 1032+2756  & c     &    103214.4+275555 &  3    \\
GRG 1032+2756  & s     &    103215.1+275115 &  1    \\
\hline
GRG 1216+4159  & c     &    21641.4+415545  &  11   \\
\hline
GRG 1343+3758  & c     &    134255.0+375819 &  2    \\
\hline
GRG 1400+3017  & n     &    140045.0+302214 &  3    \\
GRG 1400+3017  & s     &    140038.4+301325 &  3    \\
\hline
GRG 1453+3308  & n     &    145302.0+331046 &  4    \\
GRG 1453+3308  & c     &    145303.0+330856 &  2    \\
GRG 1453+3308  & s     &    145301.4+330556 &  1    \\
\hline
GRG 1521+5105  & c     &    152132.5+510232 &  7    \\
\hline
GRG 1552+2005  & c     &    155209.0+200524 &  8    \\
\hline
GRG 1738+3733  & n     &    173820.6+373658 &  2    \\
GRG 1738+3733  & c     &    173821.0+373333 &  2    \\
GRG 1738+3733  & s     &    173821.8+373108 &  1    \\
\hline
GRG 2103+6456  & c     &    210322.1+645929 &  9     \\
\hline
\end{tabular}
\end{center}
\end{table*}

\begin{table*}[!tbp]
\begin{center}
\caption{Integral flux
densities  (mJy) of the radio sources according to the data from
the RATAN-600 and the WENSS, NVSS surveys}
\begin{tabular}{r|c|c|c|c|c|c|c}
\hline
Source	    & 2.7 cm & 3.9 cm & 6.25 cm & 13 cm &  92 cm  & 21 cm & ~~6.2 cm\\
component	 & RATAN & RATAN & RATAN  & RATAN   & WENSS  & NVSS & GB6  \\
\hline
0139+3957w	& --	&  --	&  470	 &   857   &	--   &	2120  &	  656	 \\
     c	    & --    &  --   &  139   &	 252   &    --	 &   744  &   317    \\
     e	    & --    &  --   &  82    &	 212   &    --	 &   133  &   --     \\
0452+5204c	& 417	&  827	& 1141	 & 1984	   & 18760   & 3003   & 844	 \\
0751+4231c	& 103	&  227	& 274	 &  476	   & 1365    & 202    &	 35	 \\
0912+3510n	& --	&  --	& 21	 & $<$120  &	160  &	  56  & $<$20	 \\
     s	    & --    &  --   & 70     &	 144   &    512	 &   101  &    20    \\
0929+4146c	& --	&  --	& 215	 &  315	   & 1560    & 200    &	 91	 \\
1032+2759n	& --	&  --	&  46	 & $<$120  &	--   &	  92  & $<$20	 \\
     c	    & --    &  --   &  35    & $<$120  &    --	 &    75  &    59    \\
     s	    & --    &  --   &  86    &	 108   &    --	 &   138  &    56    \\
1216+4159c	& --	&  --	& 123	 &  207	   & 1604    & 264    &	 77	 \\
1400+3017n	& --	&  --	&  61	 &   178   &   1258  &	 333  &	   73	 \\
     s	    & --    &  --   &  40    &	 155   &   1053	 &   155  &    37    \\
1453+3308n	& --	&  --	&  19	 &  $<$40  &	420  &	 245  & $<$20	 \\
     c	    & --    &  --   & 109    &	 138   &    593	 &   149  &   131    \\
     s	    & --    &  --   &  76    &	$<$40  &    488	 &    89  & $<$20    \\
1521+5105c	& --	&  --	& 317	 &  549	   & 4903    & 747    & 377	 \\
1552+2005w	& --	&  --	&  82	 &   212   &	--   &	 133  & $<$20	 \\
     e	    & --    &  --   & 139    &	 252   &    --	 &   744  &   317    \\
     ee	    & --    &  --   & 470    &	 857   &    --	 &  2120  &   656    \\
1738+3733n	& --	&  --	&  36	 &    56   &	152  &	  64  & $<$20	 \\
     c	    & --    &  --   &  46    &	 113   &    720	 &   117  &    93    \\
     s	    & --    &  --   &  29    & $<$30   &    133	 &    58  & $<$20    \\
2103+6456c	& --	&  --	&$<$54	 &$<$180   &  337    & 124    &	 32	 \\
\hline
\end{tabular}
\end{center}
\end{table*}

In our recent work
\cite{r_grg1},\cite{r_grg2}, based on the
analysis of radio spectra of giant radio galaxies we have come to
the conclusion that the change in the spectral index of giant
radio galaxies, depending on the shift with respect to the
galactic center, noted earlier in
\cite{Schoenmakers_Bruyn} is linked with the
particle energy variations in the components, caused by the
pressure variation of the gas, flowing around, i.e. due to the
changes in the medium depending on the distance from the host
galaxy. However, general conclusions will be more significant at
the integral approach to the GRG population as a whole. An
essential step in the study of the causes of large sizes of giant
radio galaxies is a comparative study of similar properties of
``normal'' radio galaxies
\cite{rg_list1}-\cite{rg_list4}.
Note that  Soboleva  \cite{Soboleva}  has earlier
made the observations of radio galaxies with minute dimensions in
the centimeter wavelength range with the RATAN-600, and discovered
that the morphological structures have virtually identical
spectral indices. Therefore, the study of the GRG sample objects
will supplement the data on the radio spectra of this population
of galaxies.

\begin{table}[!tbp]
\begin{center}
\caption{Approximation
relations for the continuous radio spectra of giant radio galaxies
in the wavelength range from 92\,cm to 2.7\,cm}
\begin{tabular}{r|c}
\hline
Source	  &    Radio spectrum	 \\
component      &		   \\
\hline
0139+3957w  &  $ 3.182-1.294x	       $ \\
     c	&  $ 6.409-2.237x	   $ \\
     e	&  $ 3.135-1.182x	   $ \\
0452+5204c  &  $ 3.054-0.829x $ \\
0751+4231c  &  $ 1.971-0.697x $ \\
0912+3510n  &  $ 1.165-0.771x	       $ \\
     s	&  $ 0.922-0.560x	   $ \\
0929+4146c  &  $ 1.330-0.583x $ \\
1032+2759n  &  $ 3.366-1.400x	       $ \\
     c	&  $-0.349-0.246x	   $ \\
     s	&  $ 1.104-0.621x	   $ \\
1216+4159c  &  $ 2.288-0.882x $ \\
1400+3017n  &  $-0.914+1.406x-0.400x^2 $ \\
     s	&  $ 3.051-1.201x	   $ \\
1453+3308n  &  $ 2.602-1.126x	       $ \\
     c	&  $-1.275+12.534e^{-x}	   $ \\
     s	&  $ 1.409-0.694x	   $ \\
1552+2005w  &  $ 0.695-0.463x	       $ \\
     e	&  $34.372-19.175x+2.609x^2$ \\
     ee &  $ 3.394-0.997x	   $ \\
1521+5105c  &  $ 2.233-0.731x $ \\
1738+3733n  &  $ 0.544-0.566x	       $ \\
     c	&  $ 1.779-0.803x	   $ \\
     s	&  $ 0.607-0.561x	   $ \\
2103+6456c  &  $ 1.942-0.923x $ \\
\hline
\end{tabular}
\end{center}
\end{table}

\begin{table}[!tbp]
\begin{center}
\caption{Spectral
indices of giant radio galaxy components at 6.25 and 13\,cm}
\vspace{4mm}\begin{tabular}{r|c|c}
\hline Source  & Spectral index & Spectral index  \\
component   &	6.25 cm	  &   13 cm	 \\
\hline
0139+3957w  &	   --1.29  &  --1.29	   \\
     c	&      --2.24  &  --2.24       \\
     e	&      --1.18  &  --1.18       \\
0452+5204c  &	   --0.83  &  --0.83	   \\
0751+4231c  &	   --0.70  &  --0.70	   \\
0912+3510n  &	   --0.77  &  --0.77	   \\
     s	&      --0.56  &  --0.56       \\
0929+4146c  &	   --0.58  &  --0.58	   \\
1032+2759n  &	   --1.40  &  --1.40	   \\
     c	&      --0.24  &  --0.24       \\
     s	&      --0.62  &  --0.62       \\
1216+4159c  &	   --0.88  &  --0.88	   \\
1400+3017n  &	   --1.54  &  --1.27	   \\
     s	&      --1.20  &  --1.20       \\
1453+3308n  &	   --1.13  &  --1.13	   \\
     c	&      --0.31  &  --0.43       \\
     s	&      --0.69  &  --0.69       \\
1521+5105c  &	   --0.73  &  --0.73	   \\
1552+2005w  &	   --0.46  &  --0.46	   \\
     e	&	0.03  &	 --1.63	      \\
     ee &      --1.00  &  --1.00       \\
1738+3733n  &	   --0.57  &  --0.57	   \\
     c	&      --0.80  &  --0.80       \\
     s	&      --0.56  &  --0.56       \\
2103+6456c  &	   --0.92  &  --0.92	   \\
\hline
\end{tabular}
\end{center}
\end{table}

\begin{figure*}[!tbp]
\centerline{ \vbox{
\hbox{
 \psfig{figure=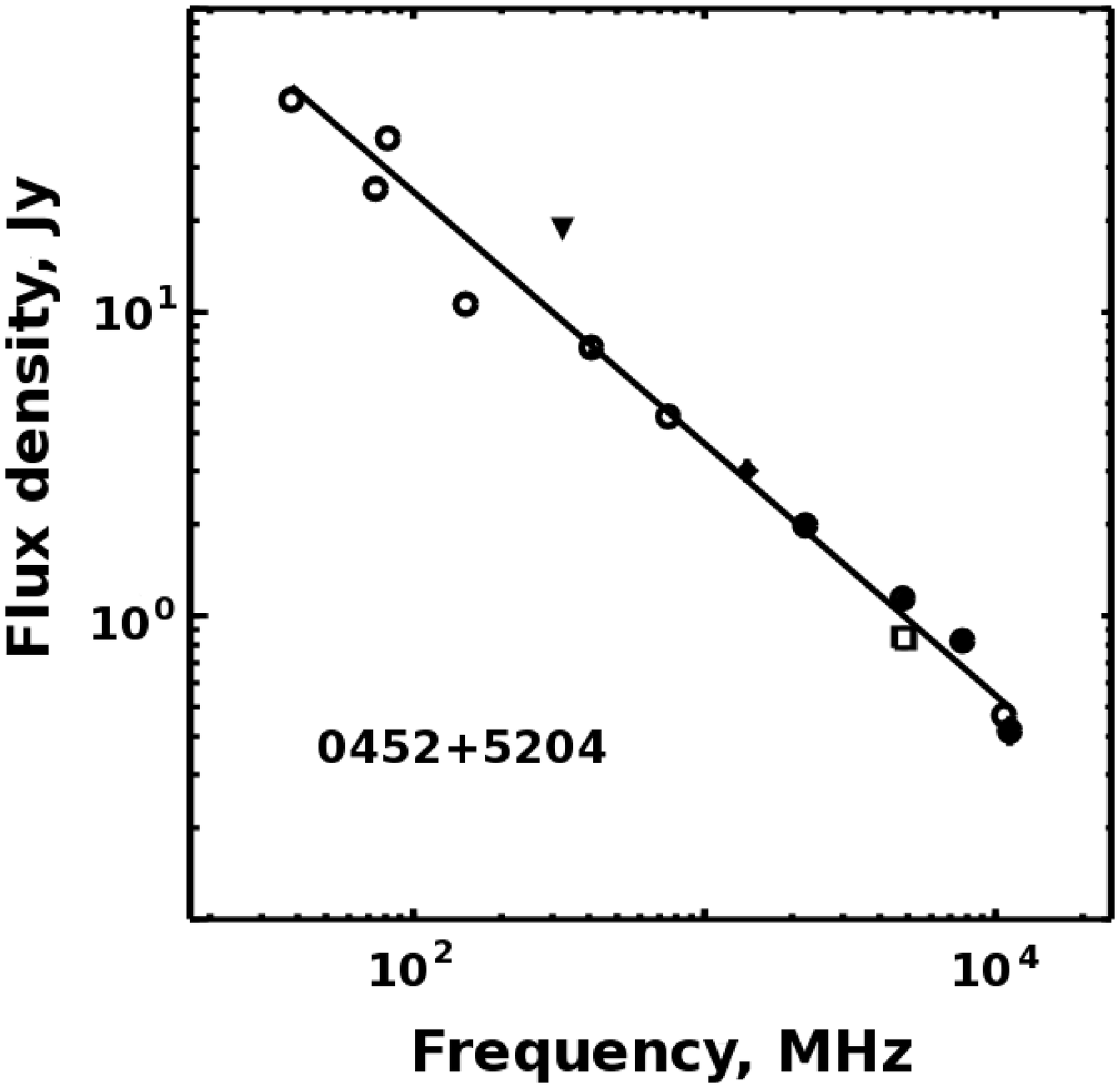,width=6cm}
\mbox{\hspace*{2cm}}
\psfig{figure=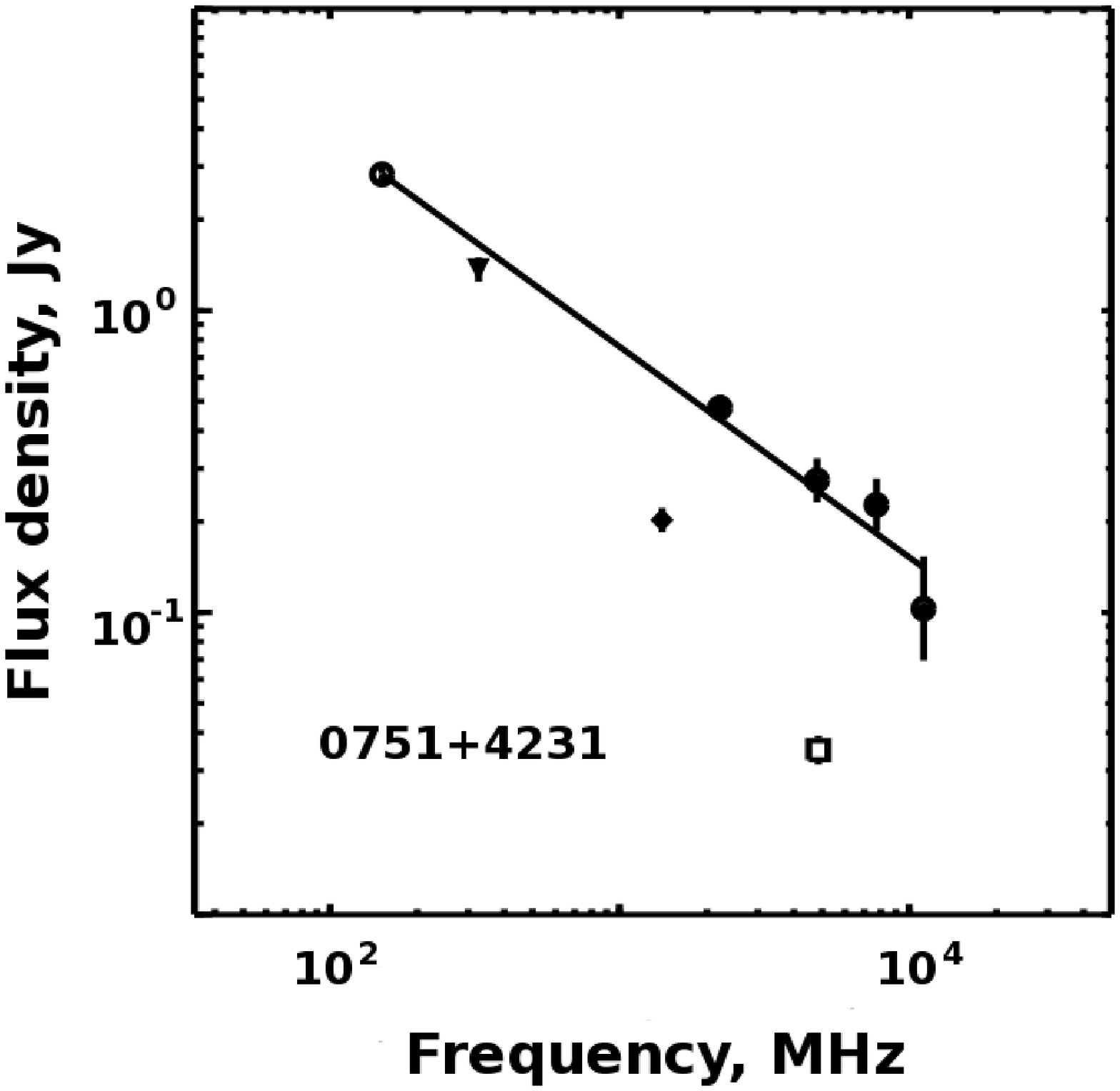,width=6cm}
} \hbox{
\psfig{figure=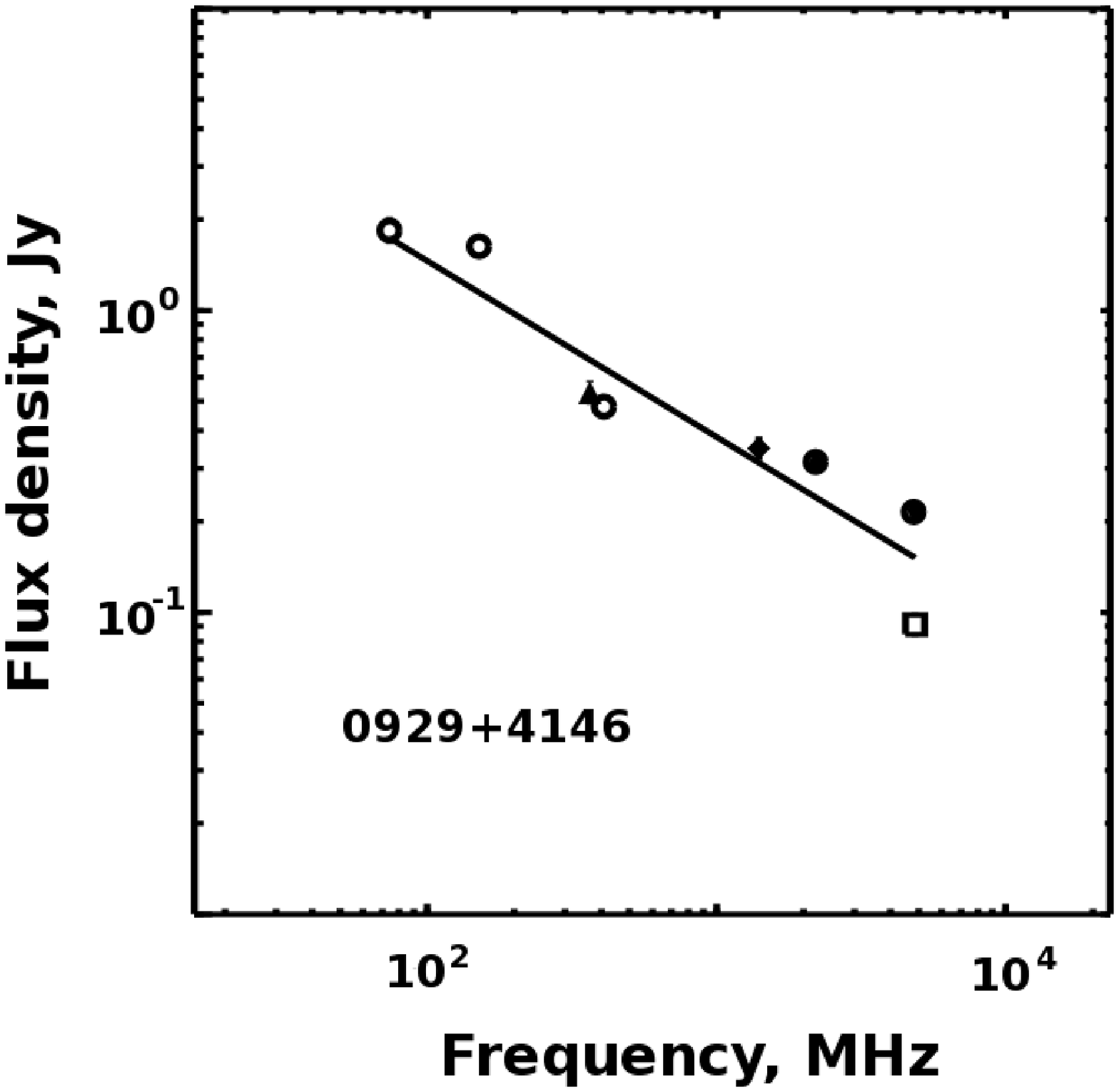,width=6cm}
\mbox{\hspace*{2cm}}
 \psfig{figure=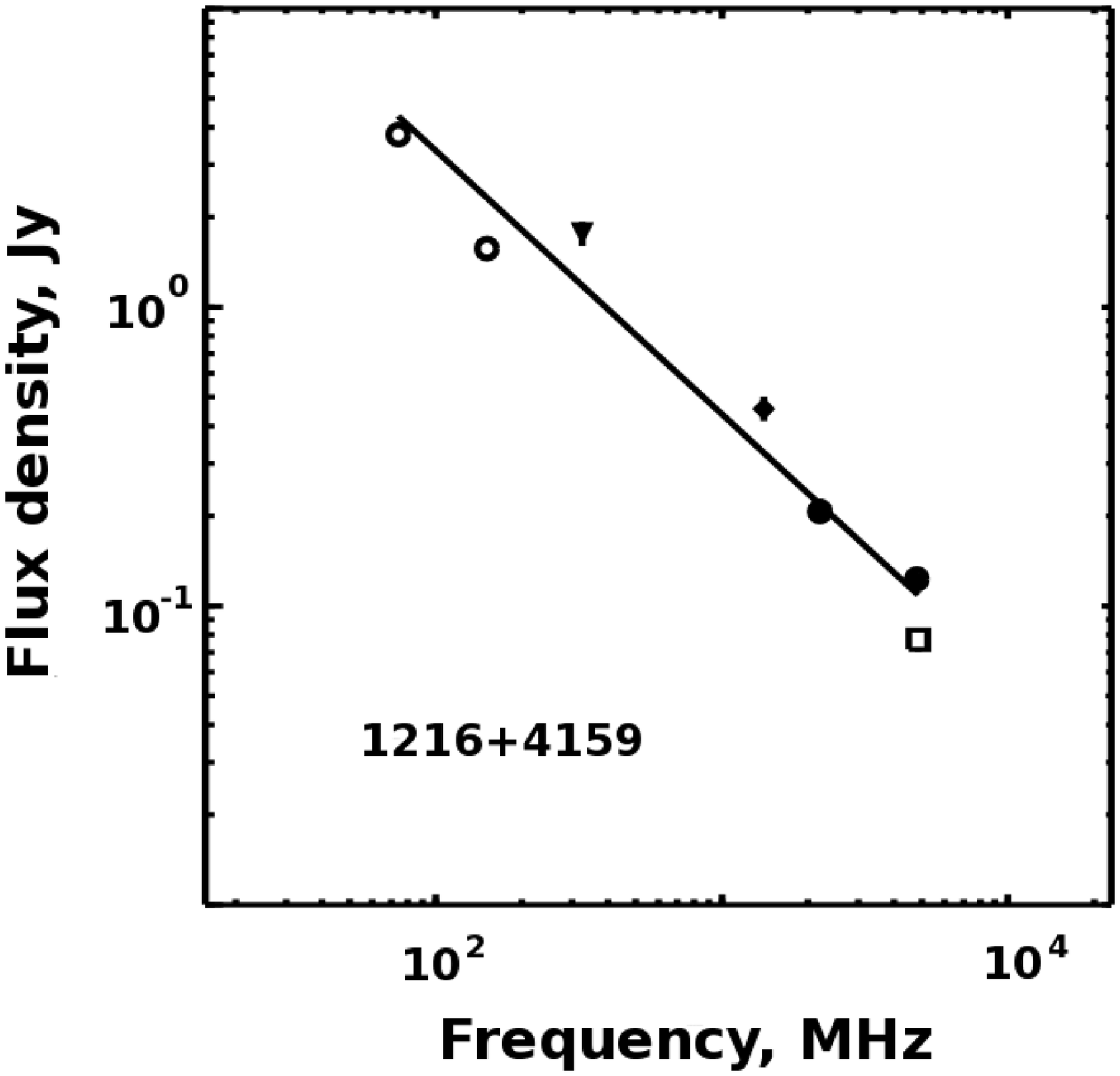,width=6cm}
} \hbox{
 \psfig{figure=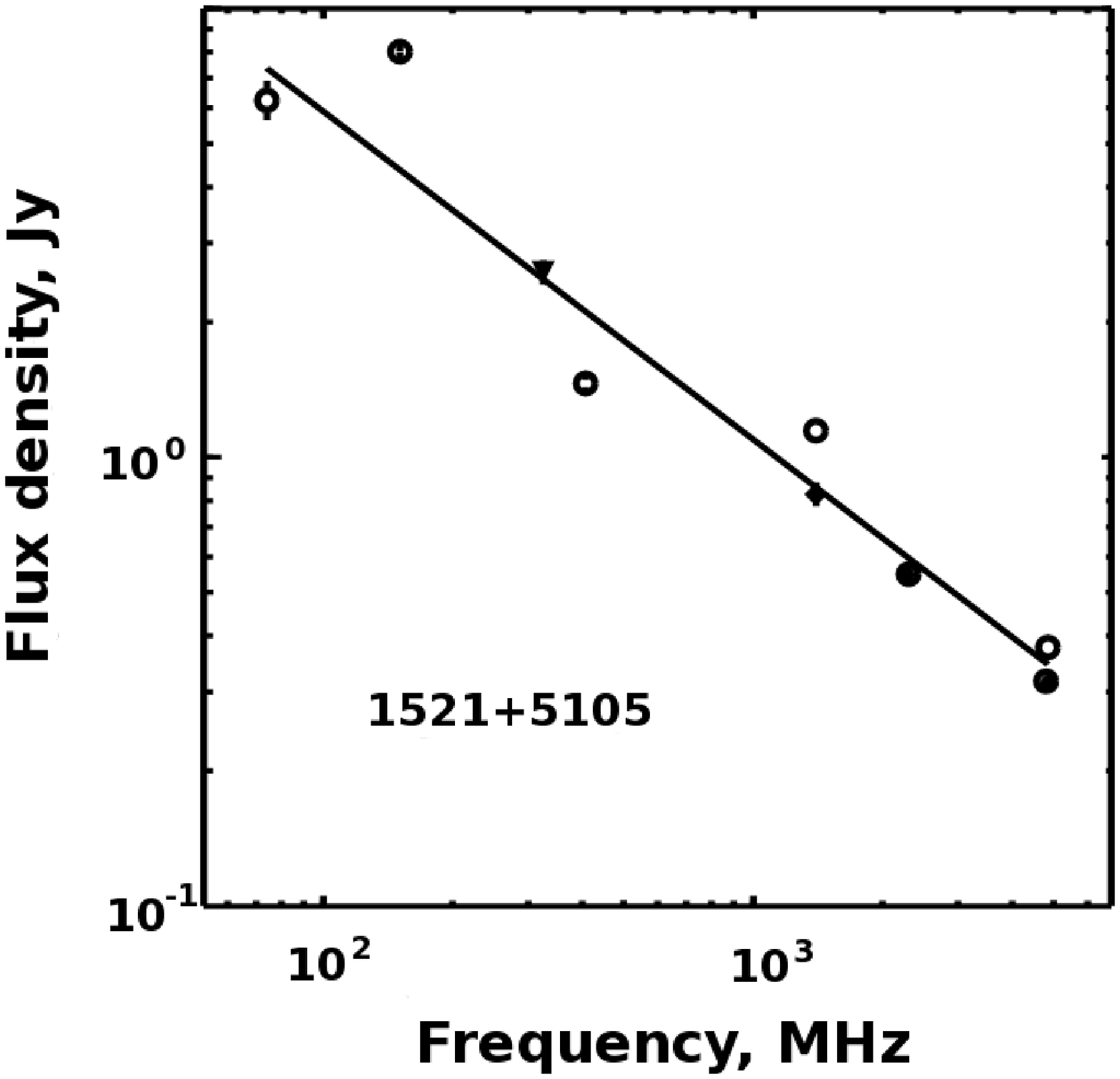,width=6cm}
\mbox{\hspace*{2cm}}
\psfig{figure=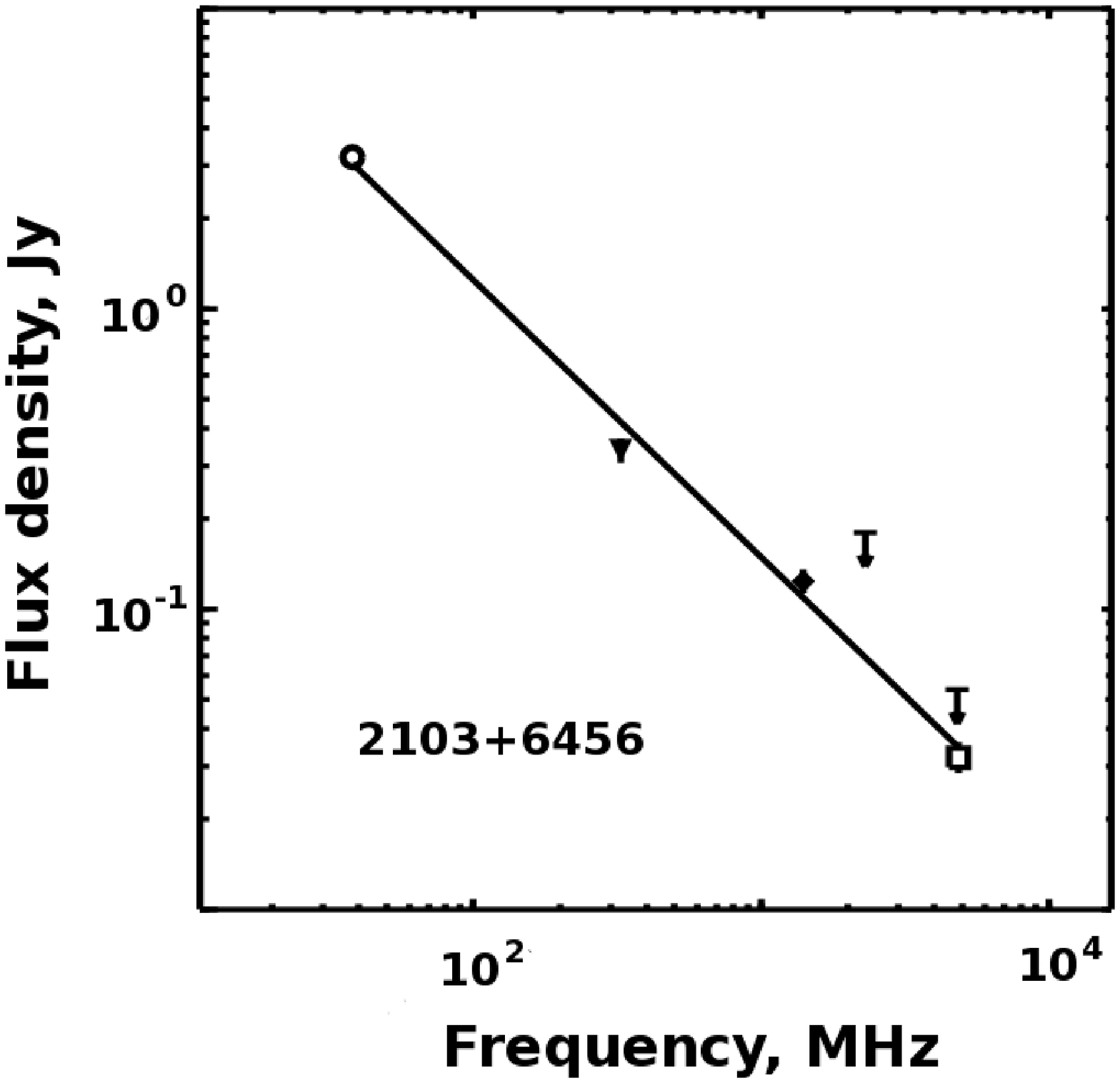,width=6cm}
} }} \caption{Radio spectra of giant radio galaxies, constructed
based on the RATAN-600 data and the NVSS, WENSS, GB6 surveys
(Table\,3) etc. The RATAN-600 data are marked by black ellipses.}
\label{ff2a:_Verkhodanov_n}
\end{figure*}

\begin{figure*}[!tbp]
 \centerline{ \vbox{
\hbox{
 \psfig{figure=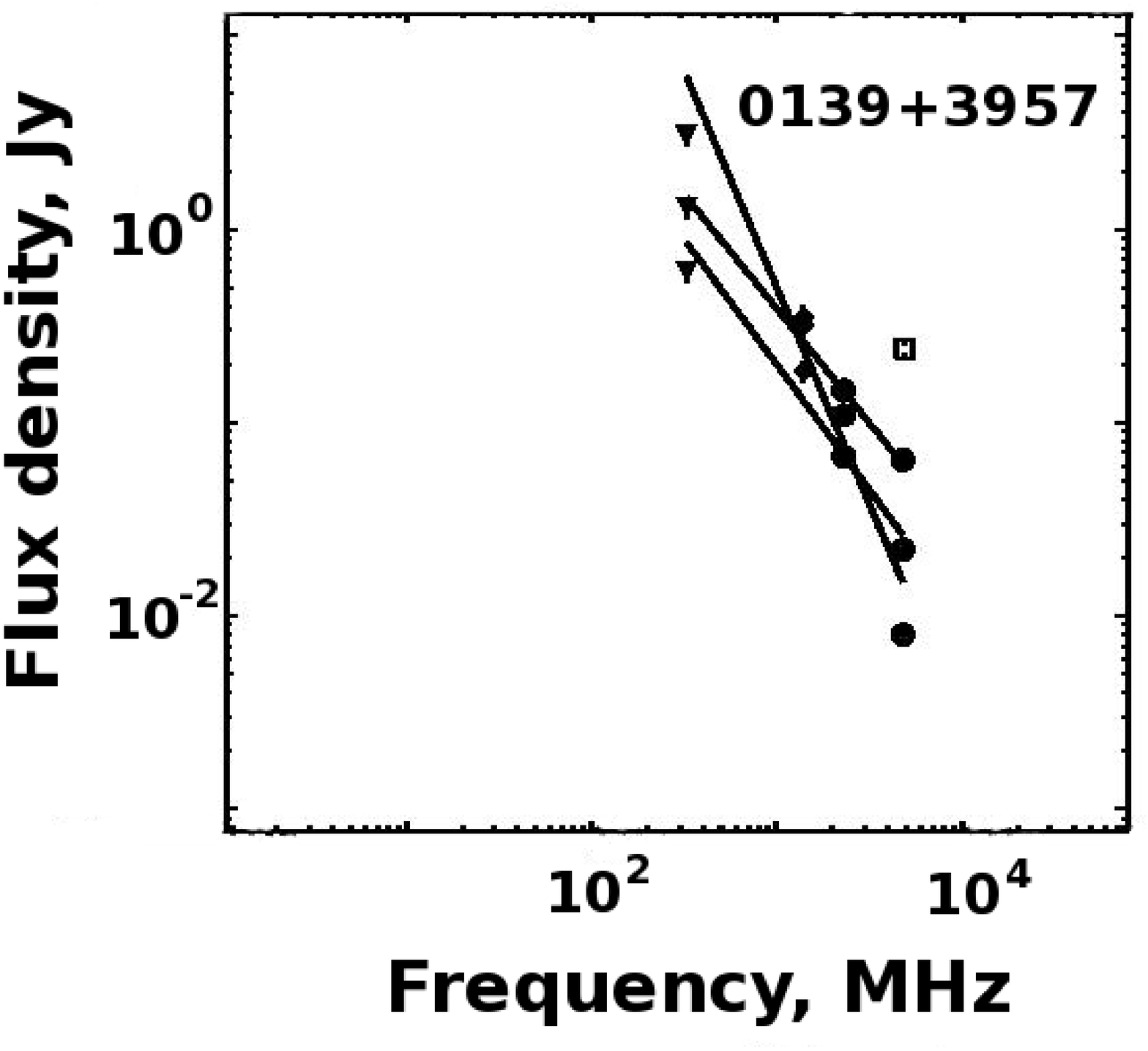,width=6cm}
\mbox{\hspace*{2cm}}
 \psfig{figure=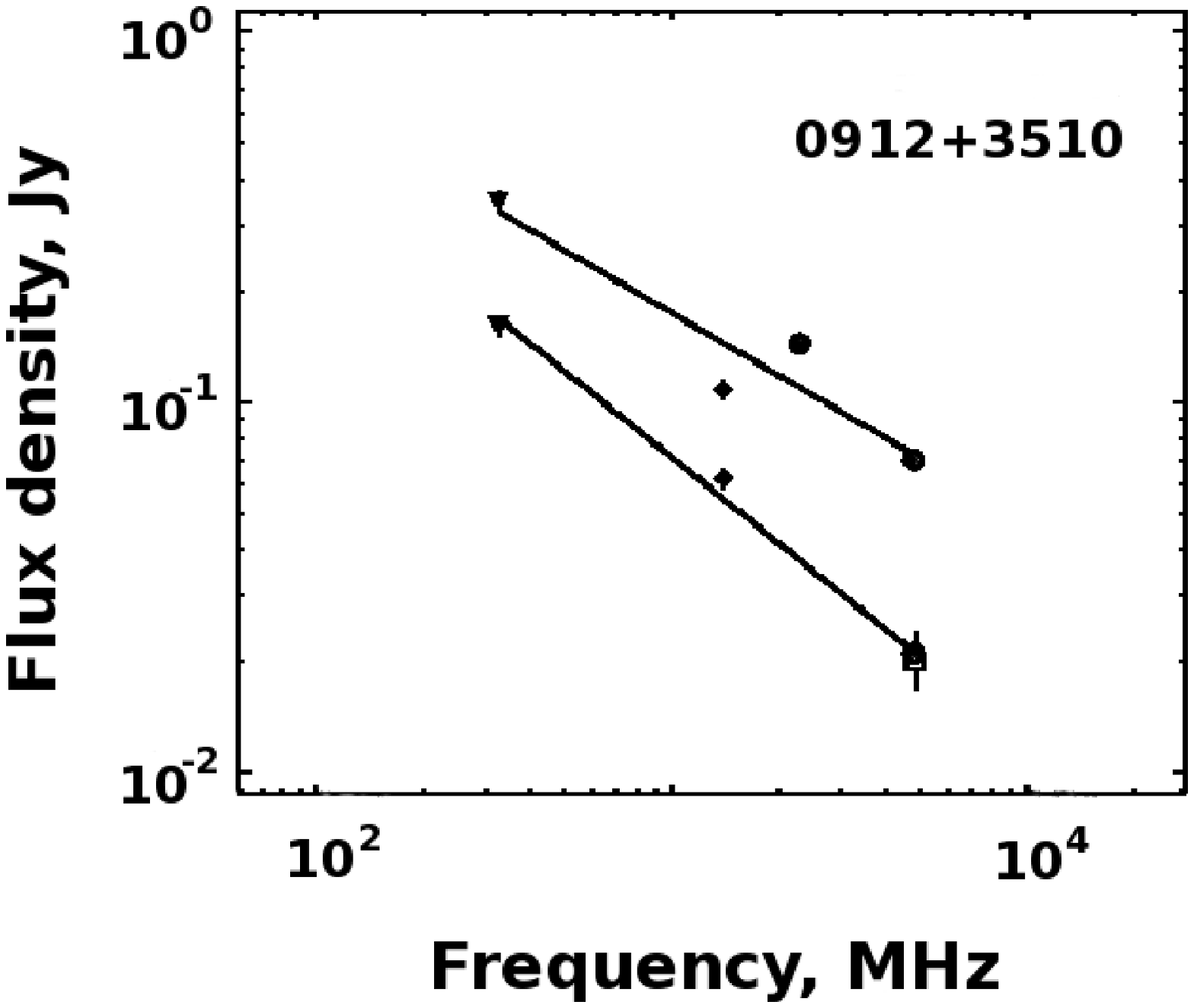,width=6cm}
} \hbox{
 \psfig{figure=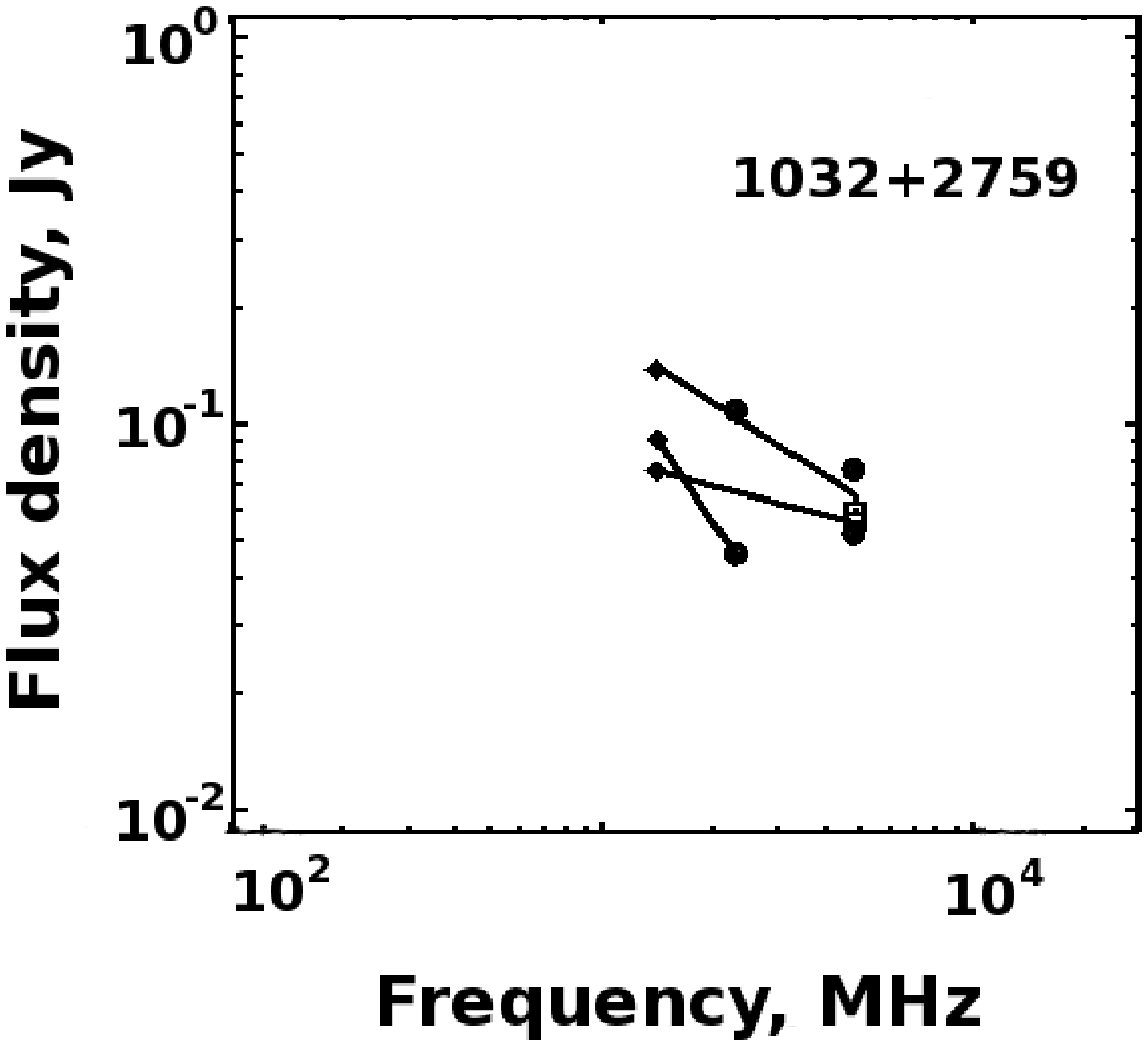,width=6cm}
\mbox{\hspace*{2cm}}
 \psfig{figure=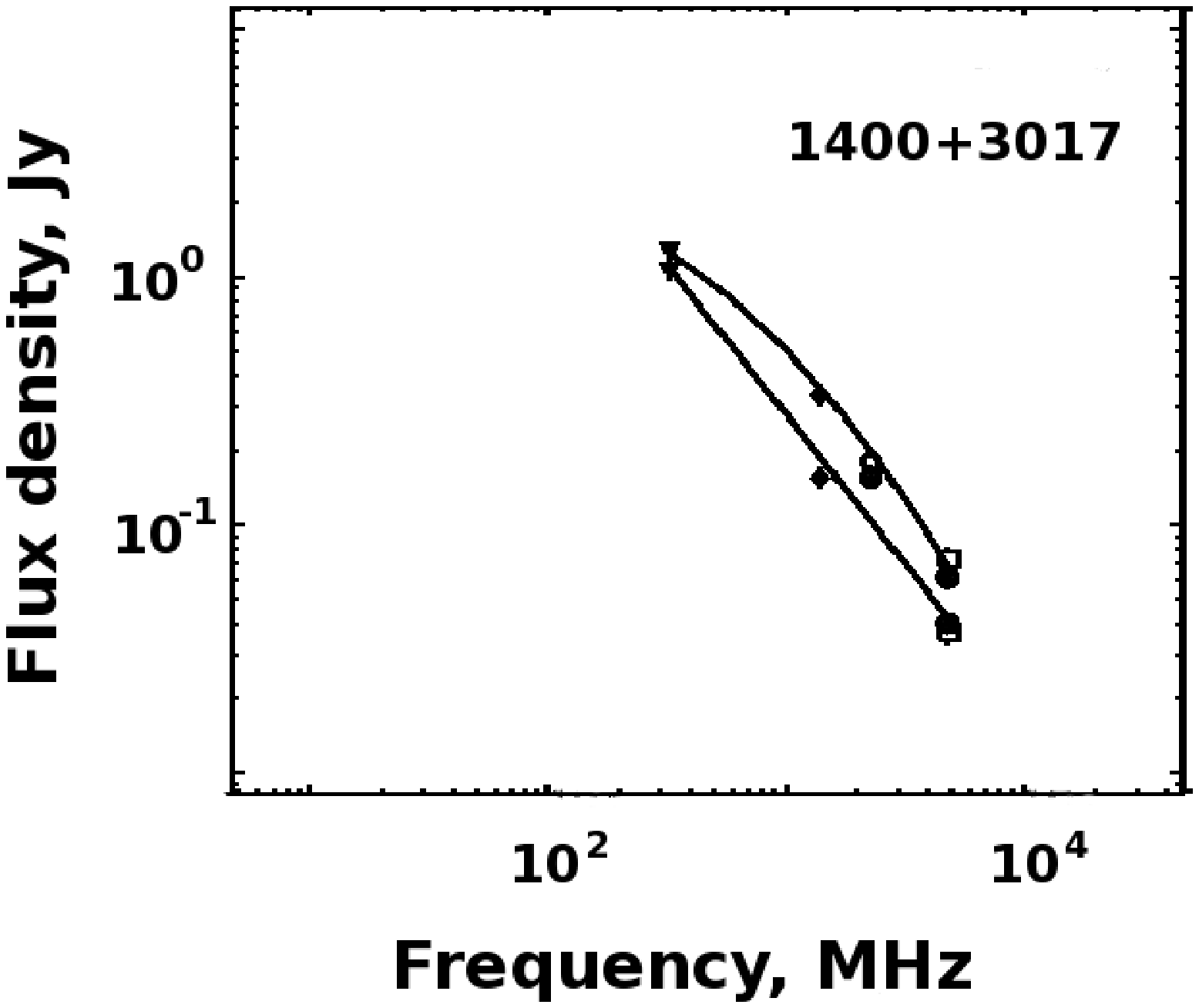,width=6cm}
} \hbox{
 \psfig{figure=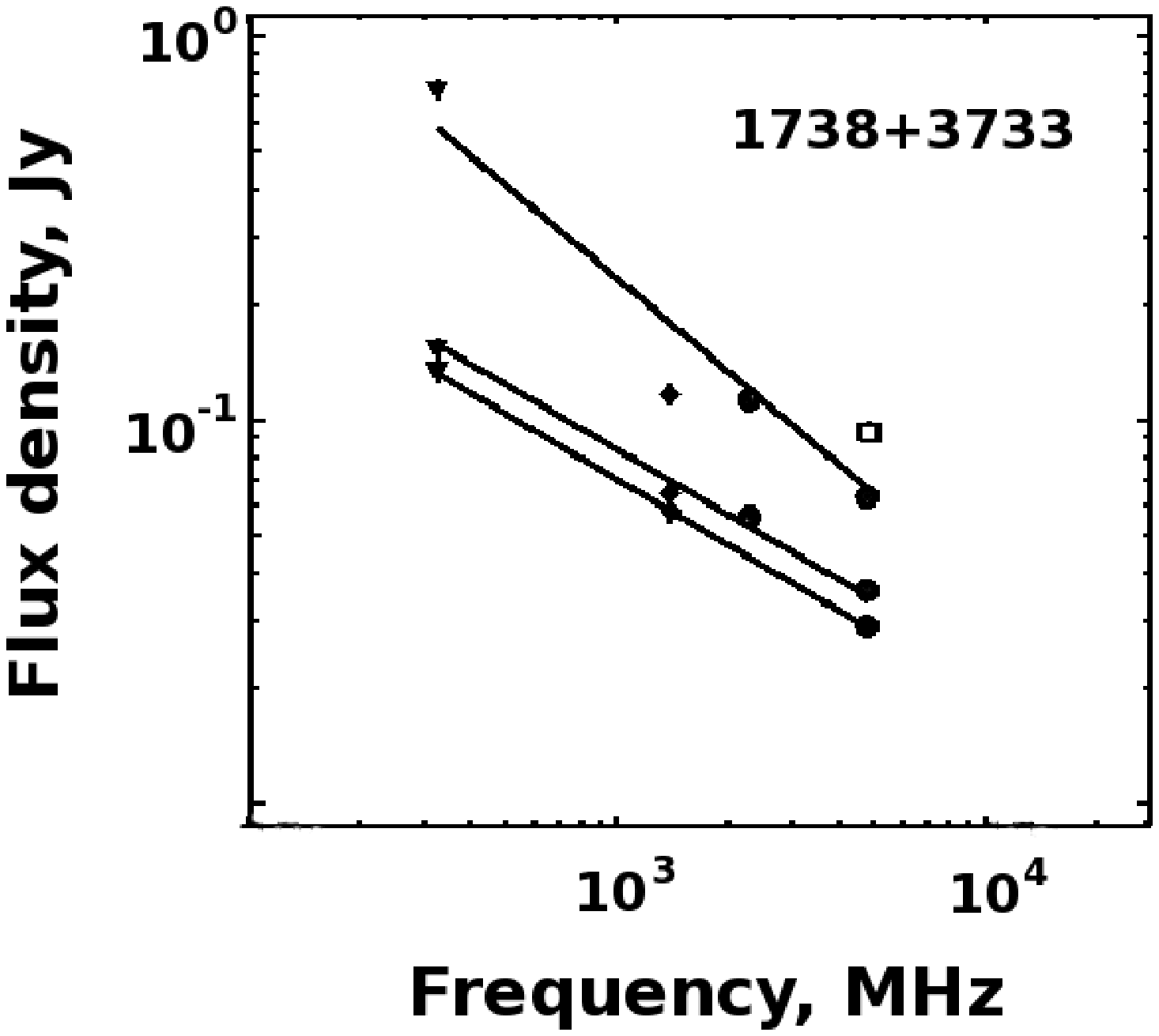,width=6cm}
\mbox{\hspace*{2cm}}
 \psfig{figure=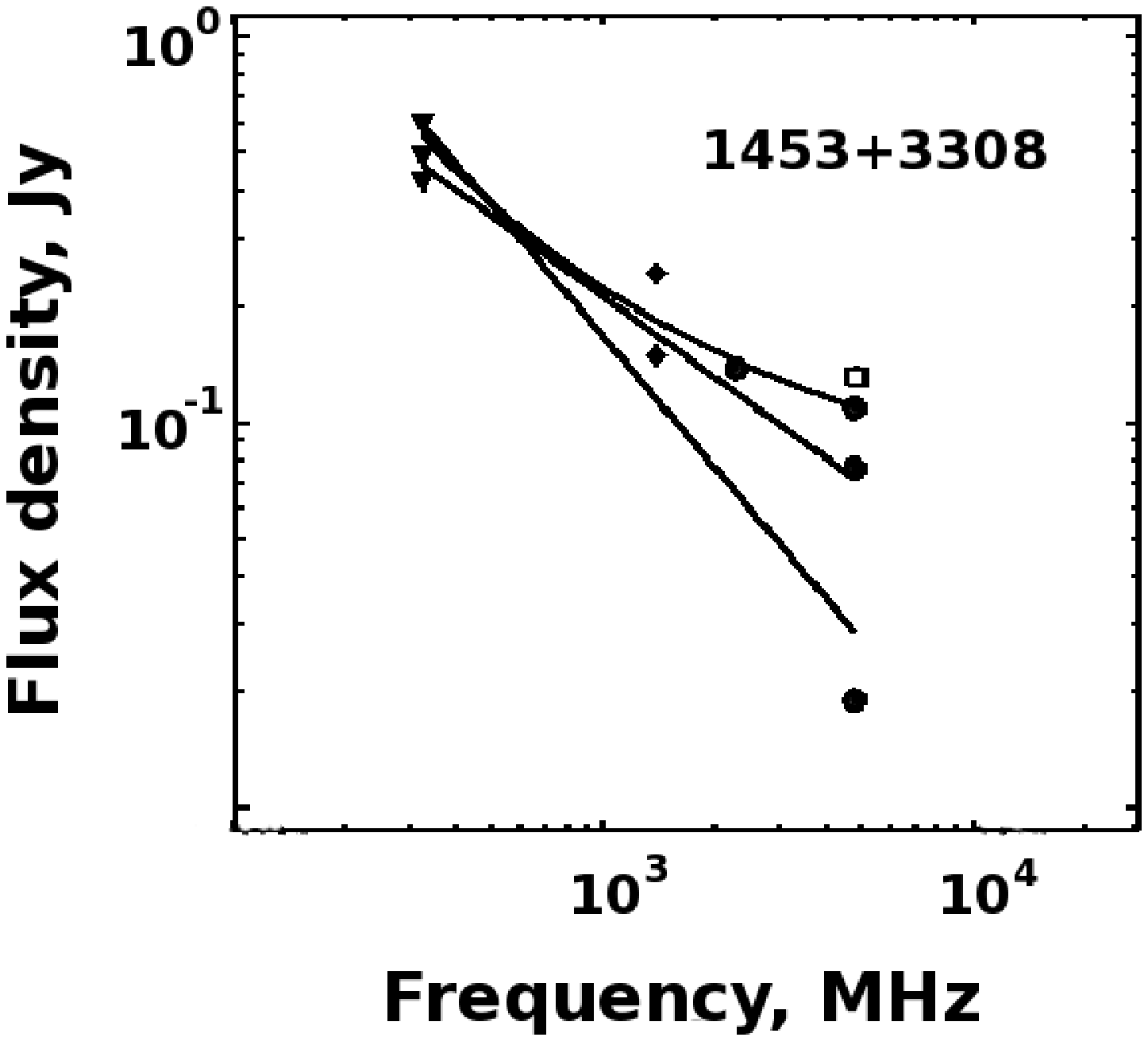,width=6cm}
} \hbox{
 \psfig{figure=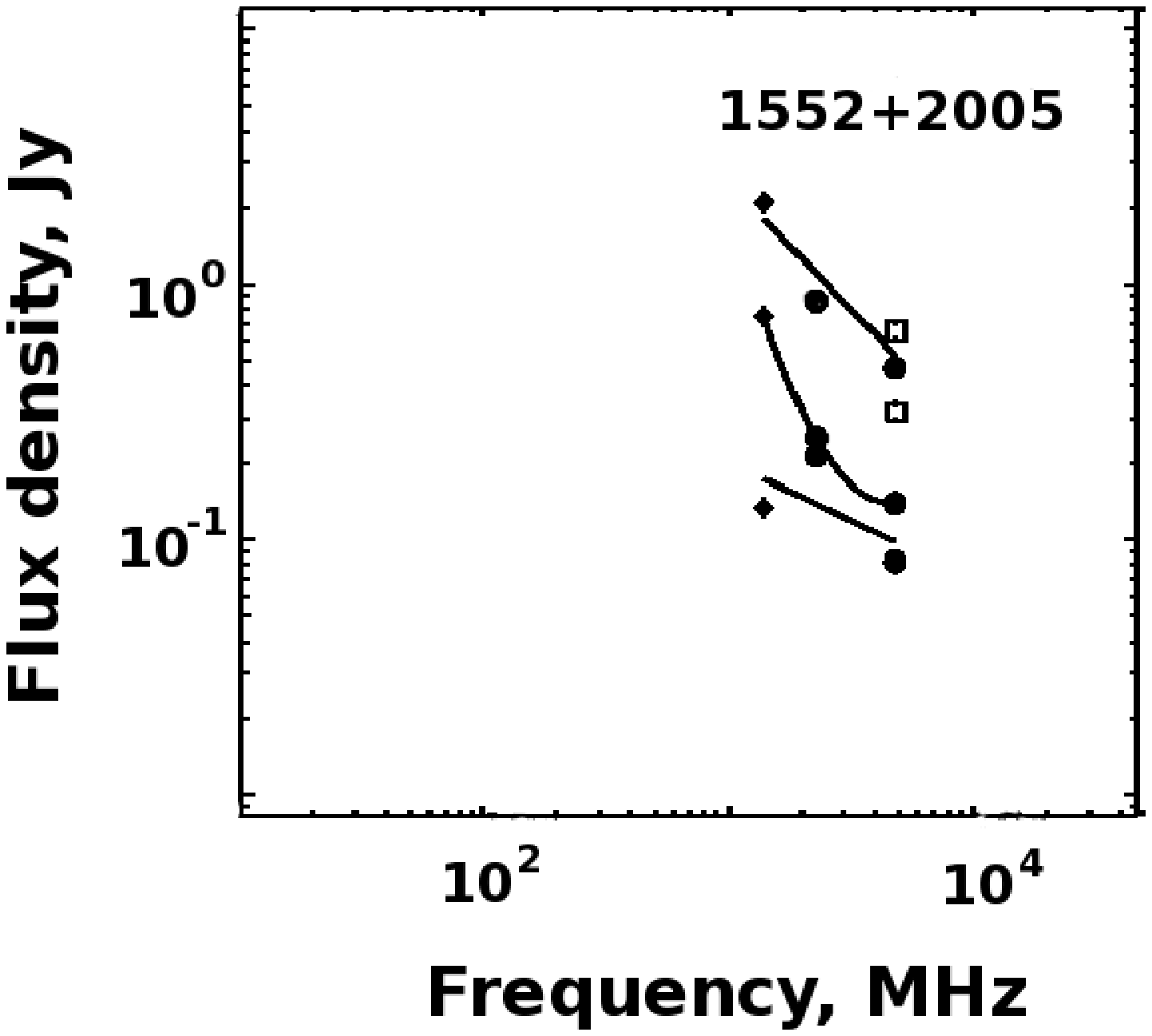,width=6cm}
} }}
{{\bf Fig.\,2.} (Contd.) }
\end{figure*}

\begin{figure}[!tbp]
\centerline{
\psfig{figure=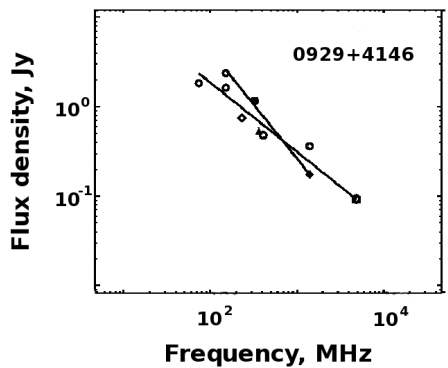,width=6cm}
}
\caption{ Radio spectra of radio galaxies: a giant GRG\,0929+4146,
having a steeper spectrum, and a normal binary J092924+414618, the
radio brightness distribution of which was integrated by the power
beam pattern in the  RATAN-600 observations. Individual spectra
for each source are shown, giving the total contribution to the
observed radio brightness distribution.}
\end{figure}

\begin{figure}[!tbp]
\centerline{
\vbox{
 \psfig{figure=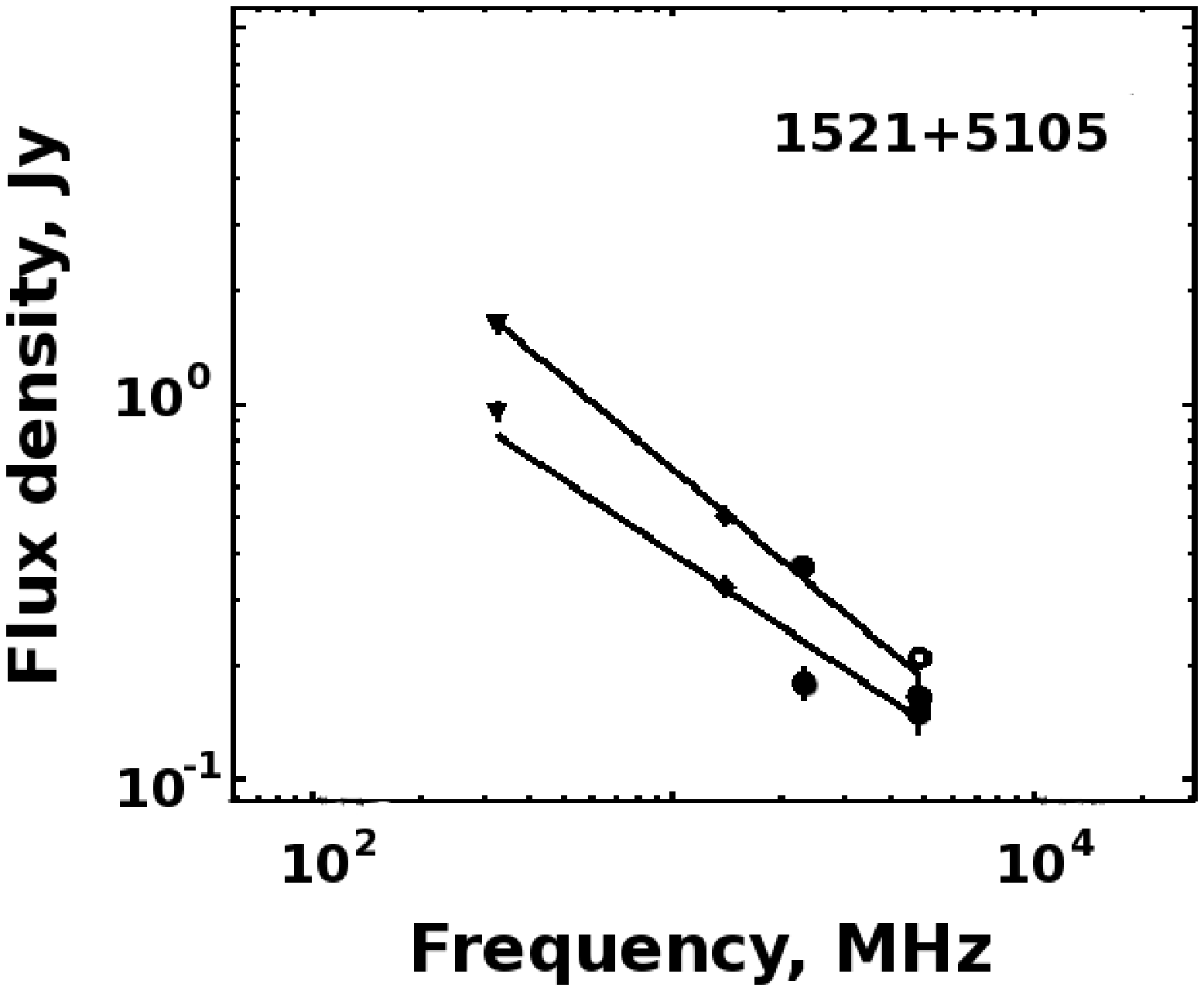,width=7cm}
 \psfig{figure=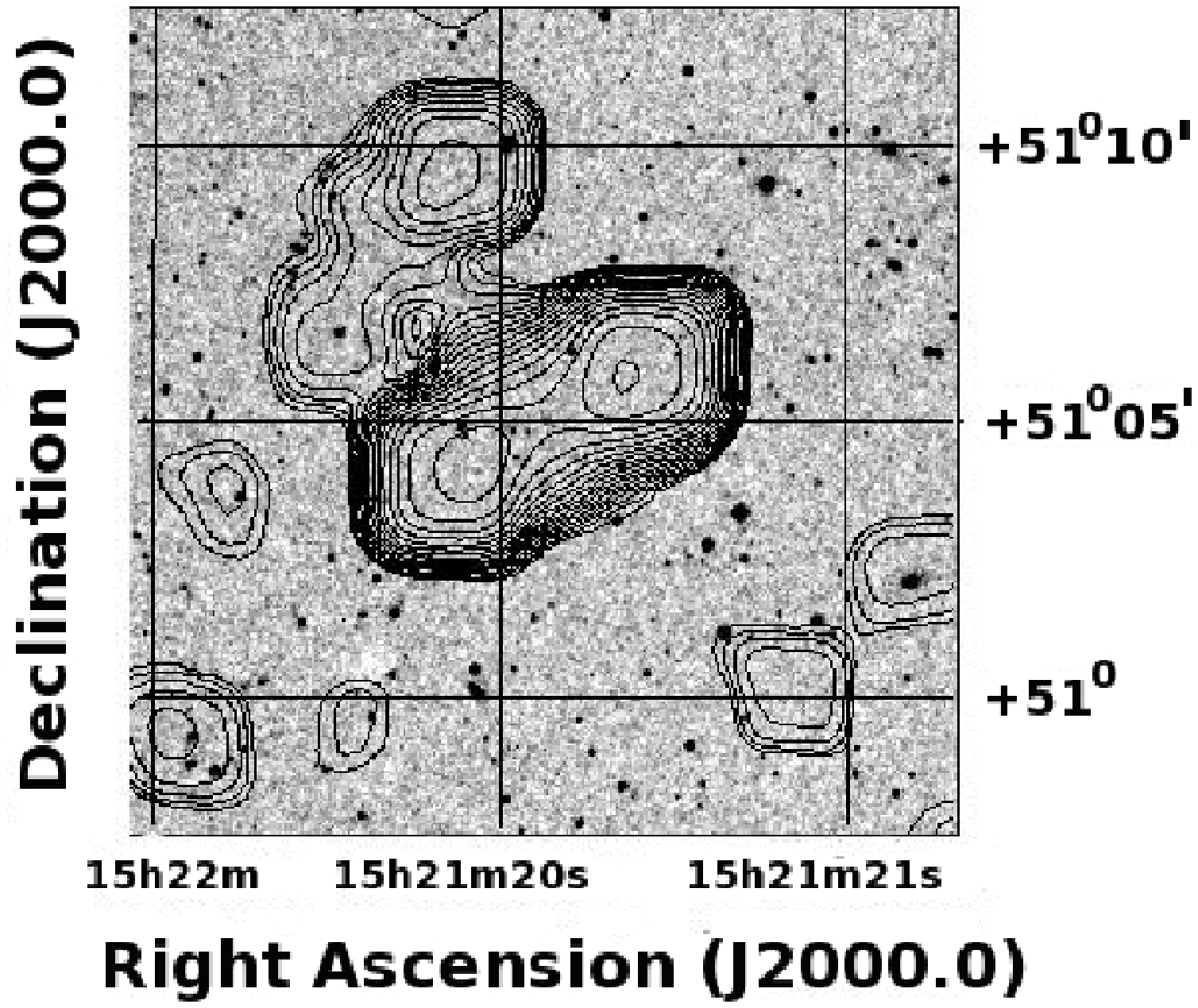,width=7cm}
\mbox{\vspace*{1cm}}
} } \caption{Top: radio spectra of the GRG\,1521+5105 components.
The RATAN-600 data area marked by black ellipses. Bottom: a radio
image of GRG\,1521+5105 (in the center) according to the NVSS
data, superimposed on the optical DSS survey image.}
\end{figure}

GRGs are as well interesting to study the contamination of the
microwave background due to their morphological properties (as
well as their size, shape and orientation) in the pixel domain, or
the phase characteristics in the harmonic domain
\cite{Verkh_Khab}. While their apparent contribution
in the background maps in the millimeter range seems to be weak,
the angular sizes of sources (up to 10 minutes of arc) produce
problems in the component separation  due to the changes in the
spectral index in the locations of extended radio components of
galaxies. Therefore, one of the interesting problems is to
evaluate and account for a possible contribution of GRGs to the
background anisotropy, both their radiation in the millimeter
range, and the effects, occurring in the component separation on
the  multipole scales \mbox{$\ell\ge500$} in different frequency
ranges.

In this paper, we present the results of measurements of the flux
densities of giant radio galaxies at the centimeter and decimeter
ranges based on the results of two observational sets at \mbox{the
RATAN-600.}

\section{RATAN-600 DATA}
\subsection{RATAN-600 Observations}

The initial GRG sample is selected based on the lists from
\cite{Schoenmakers_Mack}-\cite{Lara_Marquez}
within the operational range of \mbox{the RATAN-600.} The
observations of the GRGs were performed at the Northern sector of
the  \mbox{RATAN-600} in the second decade of December 2008, and
at the Southern sector in early January 2010. The continuous
spectrum radiometers of the Feed  Cabin 1
\cite{Nizhel} were used at the 1.38, 2.7, 3.9,
6.25, 13 and 31\,cm wavelengths. Note that despite the large range
of wavelengths, the elevated noise conditions during the period of
observations have restricted the data, suitable for the analysis
to four bands only: 2.7, 3.9, 6.25 and 13\,cm. The size of beam
patterns in the central section at the observational elevation
amounted to 25\arcsec, 36\arcsec, 43\arcsec and 57\arcsec,
90\arcsec{} and 119\arcsec{}, respectively. For the wavelengths of
6.25 and 13\,cm we used the spectral analyzer subchannels to
effectively deal with the interference. The list of observed
sources is given in Table\,1, the log book---in Table\,2. Note
that for the galaxies GRG\,1343+3758 and GRG\,2103+64 we did not
manage to reach a sufficient  signal-to-noise ratio to be able to
detect the sources.

Depending on the position angle of the radio structure, from one
to three sections of the source	 have been made	 (Table\,2). The
number of object transits through the telescope's beam pattern
(BP) was limited by the observational time, granted by the program
committee.

\subsection{Data Reduction}

To bind the flux densities with the international scale
\cite{Baars} we observed standard calibration
sources from the RATAN-600 standard list
\cite{Aliakberov},\cite{Trushkin}. The curves
of source transits were analyzed in the FADPS standard processing
system \cite{Verkh_Erukh},\cite{Verkh}. The
first stage of data processing consists in subtracting the
low-frequency trend with a window smoothing of 8 minutes of arc
from the source transit records. A transition to flux densities
was performed by integrating the extended signal, approximated by
a set of Gaussians, and a transition to the flux density scale
through calibration. The noise level in the records of single
transits at the wavelengths of 1.38, 2.7, 3.9, 6.25 and 13\,cm in
the observations at the northern sector at the elevation of
76\degr\ amounted to 8.1, 5, 36, 3.3 and 65 mK/s$^{1/2}$,
respectively, while in the southern sector at the elevation of
82\degr\ it amounted to 17.2, 8.9, 18.1, 10.7 and 96.6
mK/s$^{1/2}$, respectively. The measurements of flux densities at
the wavelengths of 2.7, 3.9, 6.25 and 13\,cm are given in
Table\,3. The table also lists the values of integrated flux
densities of the studied sources, computed from the NVSS (NRAO VLA
Sky Survey) maps \cite{Condon} at 21\,cm,
constructed on the VLA radio interferometer  (USA), and WENSS
(Westerbork Northern Sky Survey) maps
\cite{Rengelink}, constructed at the Dutch radio
interferometer in Westerbork at 92\,cm. The table as well gives
the data from the GB6 (Green Bank) catalog
\cite{Gregory}. Integration of the radio brightness
distribution in the maps was carried out using an interactive
image analysis package SkyView\footnote{{\tt
http://www.ipac.caltech.edu/Skyview/}} with a preliminary
subtraction of the trend. For identifying the objects and making
the estimates of their parameters we also used the CATS
database \cite{cats1},\cite{cats2}.
Among the CATS catalogs, we found the identifications in the GB6
 \cite{Gregory}, VLSS
\cite{Cohen}, 6C
\cite{Hales_Masson}, 7C
\cite{Riley}, 8C
\cite{Hales_Waldram}, Texas
\cite{Douglas}, and B3 \cite{Ficarra}
surveys.

One of the problems in observing the GRGs with \mbox{the
RATAN-600} is the evaluation of flux densities of multicomponent
sources, where the components have close right ascensions, but
different declination angles. In this case, the radio source is
oriented coaxially (vertically, if projected on the plane) with
the beam pattern in the observations at the meridian. Then, in
each section there appears a contribution of different radio
components of the source object due to an extended form of the
beam pattern vertically. One way to restore the obtained signal is
to simulate an extended source with more accurate data from the
NVSS catalog, its convolution with the calculated beam pattern of
\mbox{the RATAN-600} and the calculation of the signal correction,
caused by the confusion in the contributions of different
components. This approach, applied in
\cite{r_grg1} gave a somewhat inflated estimate of
the fluxes after the signal reconstruction, which may be due to
inaccurate plotting of the source shape. Therefore, we used other
estimates based on the additivity of the convolution operator. In
this case, when the same beam pattern is used to observe two
components of the source, separated by an angular distance $\Delta
h$, the observed flux density $S_1$ of one component, apart from
the incoming flux $B_1$, gets a part of the flux from the second
component $kS_2$, where $k$ is the coefficient (the value from the
antenna beam pattern), corresponding to a vertical shift from the
center of the beam by the angular distance $\Delta h$.

Hence, we find that a real signal from one component can be
estimated as  \mbox{$B_1 = (S_1 - k S_2) / (1-k^2)$.}  This
algorithm was used in the analysis of the GRG observations  in
several sections. The results are listed in Table\,3.

The error in determining the flux densities in the RATAN-600
observations of the sources with flux densities greater than
50\,mJy was about 10\%, and for the flux densities below 50\,mJy
it amounted to 13\% at the wavelength of 6.25\,cm. Similarly, at
13\,cm the error of 10\% was found for the values greater than
180\,mJy, and 15 \%---for the flux densities below 180\,mJy.

For the sources GRG\,0452+5204 and \linebreak GRG\,0751+4231 at
the wavelengths of 2.7 and 3.9\,cm the observations were carried
out in the ``beam switching'' mode. To take into account a
possible flux density drop of an extended object when observed in
this mode, we modelled the passage of two sources through two
horns. The modelling procedure included: the calculation of a
two-dimensional beam pattern of the RATAN-600 applying Korzhavin's
method \cite{Korzhavin} within the FADPS system
\cite{Verkh_Erukh} at the observed wavelength,
convolution with the components of the observed source, and
modelling the object transit through the RATAN-600 beam pattern.
In addition, we also used in the data analysis the BP calculations
made by Majorova \cite{Majorova}. The conversion
factor estimate of the integral flux density of a given extended
radio source in the beam switching mode model was accounted for in
the analysis of real observations.

\subsection{Spectra}

Based on the  measurement data we have constructed the spectra of
the radio source components. Describing the spectra, we fitted
them by the formula $ \lg\,S(\nu) = A + Bx + C f(x)$, where $S$ is
the flux density in Jy, $x$ is the frequency  logarithm $\nu$ in
MHz, and $f(x)$ is one of the following functions: $\exp(-x)$,
$\exp(x)$ or $x^2$. The {\it spg} system was used for the spectral
analysis \cite{Verkh_radio}. The spectra of the
components are demonstrated in Fig.\,2. The analytical description
of the continuous spectrum curves for the components of the
studied GRGs is shown in Table\,4.

\section{DISCUSSION OF RESULTS}

The spectra constructed (Fig.\,2) demonstrate a variety of GRG
properties. The fact that the spectral indices in the components
of the observed radio galaxies vary significantly is obvious, even
the shapes of the spectra are different: from a very steep
spectrum of the GRG\,0139+3957 source, to the spectra with
flattening, like in the GRG\,1453+3308 source component.

For the sources, observed only in the central section, the values
of flux densities, measured at the wavelengths of 2.7, 3.9, 6.25
and 13\,cm, are listed in Table\,3, and spectral indices amount to
the $x$ argument coefficients from Table\,4. For the rest of
sources the results are presented in Table\,5. Note that in the
case of the GRG\,0751+4231 source spectrum the RATAN-600 data
points are located higher than the data from the GB6 and NVSS. In
the GB6 catalog this object is nearly a point object, which
explains the low level of the cataloged flux value. It is most
likely that the level of the corresponding value from the NVSS is
caused by the same factor as in the case when the integration by
peak values leads to an incomplete account of weak diffuse
emission. The object GRG\,1738+3733 stands out among the observed
sources, as both of its extended components have similar radio
spectra, and identical spectral indices.

Note that the change in the spectral index of giant radio
galaxies, depending on the shift from the galactic center was
already noted \cite{Schoenmakers_Bruyn}. It
is associated with particle energy variations in the components,
caused by the pressure variation in the surrounding gas, i.e. it
is due to the changes in the medium, depending on the distance
from the host galaxy.

The radio source, observed on the RATAN-600  in the region of
GRG\,0929+4146 turned out to be composed of two radio galaxies:
GRG\,0929+4146 as such in the shape of a multidimensional object,
stretching along one line, and a double type-FR\,II radio source
with the coordinates ($\alpha=09^h29^m24^s$,
\mbox{$\delta=+41\degr46'18''$}), which merge into one extended
object that can be seen to the left of the GRG\,0929+4146 in
Fig.\,1. The RATAN-600 does not resolve radio galaxies in the
meridian transits, and hence Fig.\,2 presents the total spectrum.
We built a separate spectrum for each radio galaxy according to
the NVSS, WENSS and 7C survey data (see Fig.\,3). The integral
radio spectrum of GRG\,0929+4146 alone was approximated by the
dependence $y=3.046-1.208x$, and the spectrum of the neighboring
radio galaxy was approximated by the dependence $y=1.818-0.774x$,
which has a smaller slope and thus demonstrates that the
short-wave observations on the RATAN-600 are dominated by the
radio emission from the source	J092924+414618.

The radio galaxy GRG\,1521+5105 was resolved into two components.
Their spectra are demonstrated in Fig.\,4. The integrated radio
emission flux densities for the two components are: 368 mJy at the
wavelength of 13\,cm and 167 mJy at 6.25\,cm for J152103+510600,
and 181 mJy and \mbox{150 mJy}, respectively for J152125+510401.

The approximations for the radio spectra of the components are
described by the functions:  \linebreak \mbox{$y=2.226-0.800x$}
and $y=1.537-0.645x$. The radio galaxy	 GRG\,1521+5105,
identified with the galaxy SDSS\,J152114.55+510500.9, and having
the photometric redshift of $z=0.37$ (according to the NED
database\footnote{\tt http://nedwww.ipac.caltech.edu}), is located
on the outskirts of the projection of the cluster
NSCS\,J152018+505306 with the redshift	$z=0.52$ (NED) at the
angular distance of 15 arc minutes from the center. Nevertheless,
within 10 minutes of arc from the radio galaxy there are more than
1700 galaxies (according to the NED), and a large number of radio
sources (Fig.\,4). As we do not have any redshift data for the
vast majority of them it is difficult to judge the physical
belonging of GRG\,1521+5105 to any group of galaxies. Still, a
rich neighborhood of this radio galaxy brings additional interest
to search for the reasons of its gigantic size.

The RATAN-600 observations allowed to specify the GRG component
spectra and estimate their fluxes in the millimeter wavelength
range at the extrapolation of the integrated radio spectrum. The
flux densities of the studied GRG components lie in this part of
the spectrum at the level above 0.6\,mJy. As the expected number
of GRG-type objects amounts to several hundred on the full sphere
\cite{Verkh_Khab}, their contribution to the
background radiation can, in principle,	 result in a bias in
computing the background fluctuation level, not to mention the
problem of signal isolation.

We plan to further accumulate the data, compile the lists of new
GRGs and perform their observations with the RATAN-600.

\noindent
{\small

\section{ACKNOWLEDGMENTS}
The authors thank Yu.~Sotnikova for her help with the RATAN-600
observations and S.A.~Trushkin for valuable discussions. The study
made use of the NED database of extragalactic objects, the
NASA/IPAC Extragalactic Database, operated by the Jet Propulsion
Laboratory, California Institute of Technology, under the contract
with the National Aeronautics and Space Administration. The
authors also used the CATS\footnote{\tt http://cats.sao.ru}
\cite{cats1},\cite{cats2} database of radio
astronomy catalogs, and the FADPS\footnote{\tt
http://sed.sao.ru/$\sim$vo/fadps\_e.html}
\cite{Verkh_Erukh},\cite{Verkh} system for
processing the radio astronomy data. This work was supported by
the Leading Scientific Schools of Russia (S.\,M.\,Khaikin school)
grant, and the RFBR grant (project nos. 09-02-92659-IND and
09-02-00298). O.V.V. is also grateful for the partial support of
the Dynasty foundation.


\begin{thebibliography}{}

\bibitem{Fanaroff}
B.~L.~Fanaroff and  J.~M.~Riley, MNRAS~ {\bf 167}, 31 (1974).

\bibitem{Strom}
 R.~G.~Strom and  A.~G.~Willis, A\&A~ {\bf 85}, 36  (1980).

\bibitem{Kaiser_Dennett}
C.R.~Kaiser, J.~Dennett Thorpe, and P.~Alexander, MNRAS~ {\bf
292}, 723 (1997).


\bibitem{Blundell}
K.~Blundell, S.~Rawlings, and C.J.~Willott,
 AJ~  {\bf 117}, 677 (1999).

\bibitem{Kaiser_Alexander}
C.R.~Kaiser and P.~Alexander,
 MNRAS~  {\bf 302}, 515 (1999).

\bibitem{Komberg}
B.~L.~Komberg, I.~N.~Paschenko, Astronomy Reports~ {\bf 53}, 1086
(2009),
	 arXiv:0901.3721.
\bibitem{Mack}
K.~H.~Mack, U.~Klein, C.~P.~O'Dea, et al. 
  A\&A~ {\bf 329}, 431 (1998).

\bibitem{Jamrozy_Machalski}
M.~Jamrozy, J.~Machalski,  K.~H.~Mack, and U.~Klein,
   A\&A~ {\bf 433}, 467 (2005).

\bibitem{Schoenmakers_Mack}
 A.~P.~Schoenmakers, K.~H.~Mack,  A.~G.~de Bruyn, et al.,
    A\&AS {\bf 146}, 293 (2000).

\bibitem{Schoenmakers_Bruyn}
 A.~P.~Schoenmakers,  A.~G.~de Bruyn, H.~J.~A.~Roettgering, and
H.~ van der Laan,
    A\&A~ {\bf 374}, 861 (2001).

\bibitem{Lara_Marquez}
L.~Lara, I.~Marquez, W.~D.~Cotton, et al.,
 A\&A~ {\bf 378}, 826 (2001).

\bibitem{Lara_Giovannini}
L.~Lara, G.~Giovannini, W.~D.~Cotton, et al.,
 A\&A~ {\bf 421}, 899  (2004).


\bibitem{Saripalli}
L.~Saripalli,  R.~W.~Hunstead, R.~Subrahmanyan, and E.~Boyce,
     AJ~ {\bf 130}, 896 (2005).



\bibitem{Konar}
C.~Konar, D.~J.~Saikia,	 C.~H.~Ishwara-Chandra, and
V.~K.~Kulkarni, MNRAS~ {\bf 355}, 845 (2004).

\bibitem{Konar_Jamrozy}
C.~Konar, M.~Jamrozy, D.~J.~Saikia, and J.~Machalski, MNRAS~ {\bf
383}, 525 (2008).

\bibitem{Jamrozy_Konar}
M.~Jamrozy, C.~Konar, J.~Machalski, and D.~J.~Saikia, MNRAS~ {\bf
383}, 525 (2008).

\bibitem{Machalski}
 J.~Machalski, M.~Jamrozy, S.~Zola, and D.~Koziel, A\&A~
       {\bf 454}, 85 (2006).

\bibitem{Nandi}
S.~Nandi, A.~Pirya, S.~Pal,  et al.,
MNRAS~ {\bf 404}, 433 (2010), arXiv:1001.3998.

\bibitem{r_grg1}
M.~L.~Khabibullina, O.~V.~Verkhodanov, M.~Singh et al., Astron. Zh.~ {\bf
87}, 627 (2010), arXiv:1009.4539.


\bibitem{r_grg2}
M.~L.~Khabibullina, O.~V.~Verkhodanov, M.~Singh et al., Astron.
Rep. {\bf 55}, 392 (2011), arXiv:1108.3295.

\bibitem{rg_list1}
 M.~L.~Khabibullina and O.~V.~Verkhodanov,
   Astrophys. J. Suppl.~ {\bf 64}, 123 (2009), arXiv:0911.3741.

\bibitem{rg_list2}
 M.~L.~Khabibullina and O.~V.~Verkhodanov,
    Astrophys. J. Suppl.~ {\bf 64}, 276 (2009), arXiv:0911.3747.

\bibitem{rg_list3}
 M.~L.~Khabibullina and O.~V.~Verkhodanov,
   Astrophys. J. Suppl.~ {\bf 64}, 340 (2009), arXiv:0911.3752.

\bibitem{rg_list4}
O.~V.~Verkhodanov and M.~L.~Khabibullina, Astron. Lett. {\bf 36},
7 (2010), arXiv:1003.0577.


\bibitem{Soboleva}
N.~S.~Soboleva, Astrofiz. Issl. (Izvestiya SAO)	 {\bf 14}, 50
(1981).

\bibitem{Verkh_Khab}
   O.~V.~Verkhodanov, M.~L.~Khabibullina, M.~Singh, et al.,
in {\it Proc. Intern. Conf. Problems of Practical Cosmology},
Ed. by Yu.V. Baryshev, I.N.Taganov, and P. Teerikorpi (Russian
Geograph. Soc., St. Petersburg, 2008), 
p. 247.

\bibitem{Nizhel}
N.~A.~Nizhelskii, A.~B.~Berlin, A.~M.~Pilipenko et al.,
in {\it Proc. All-Russian Astroph. Conf. VAK-2001} (St.
Petersburg, 2001), p.133.

\bibitem{Baars}
J.~W.~M.~Baars, R.~Genzel,  I.~I.~K.~Pauliny-Toth, and A.~Witzel,
 A\&A~ {\bf 61}, 99 (1977).

\bibitem{Aliakberov}
  K.~D.~Aliakberov et al.,
  Astrofiz. Issl. (Izvestiya SAO)
    {\bf 19}, 60 (1985).

\bibitem{Trushkin}
S.~A.~Trushkin, {\it Spravochnik Nabliydatelia v Radiokontinyyme
(Observer's Guide in the Radio Continuum)}, {\tt
http://w0.sao.ru/hq/lran/ \linebreak /manuals/ratan\_manual.html}
[in Russian].

\bibitem{Verkh_Erukh}
O.~V.~Verkhodanov, B.~L.~Erukhimov, M.~L.~Monosov, et al., 
Bull. of the SAO {\bf 36}, 132 (1993).


\bibitem{Verkh}
O.~V.~Verkhodanov,
     ASP Conf. Ser., {\bf 125}, 46 (1997).

\bibitem{Condon}
 J.~J.~Condon, W.~D.~Cotton,  E.~W.~Greisen, et al.,
 AJ~ {\bf 115}, 1693 (1998).

\bibitem{Rengelink}
  R.~B.~Rengelink,  Y.~Tang, A.~G.~de Bruyn,  et al.,
    A\&AS {\bf 124}, 259 (1997).

\bibitem{Gregory}
 P.~C.~Gregory,	 W.~K.~Scott,  K.~Douglas, and	J.~J.~Condon,
     ApJS {\bf 103}, 427 (1996).

\bibitem{cats1}
O.~V.~Verkhodanov, S.~A.~Trushkin, H.~Andernach and V.~N.~Chernenkov,
     Bull. of the SAO {\bf 58}, 118 (2005), arXiv:0705.2959.

\bibitem{cats2}
O.~V.~Verkhodanov, S.~A.~Trushkin, H.~Andernach, and
V.~N.~Chernenkov,
     Data Science Journal {\bf 8}, 34 (2009), arXiv:0901.3118.

\bibitem{Cohen}
 A.~S.~Cohen, W.~M.~Lane, W.~D.~Cotton, et al.,
AJ~ {\bf 134}, 1245 (2007).

\bibitem{Hales_Masson}
 S.~E.~G.~Hales, C.~R.~Masson, P.~Warner, et al.,
MNRAS~ {\bf 262}, 1057 (1993).

\bibitem{Riley}
J.~M.~W.~Riley, E.~M.~Waldram, and J.~M.~Riley, 
MNRAS~ {\bf 306}, 31 (1999).

\bibitem{Hales_Waldram}
S.~E.~G.~Hales, E.~M.~Waldram, N.~Rees, and P.~J.~Warner, 
MNRAS~ {\bf 274}, 447 (1995).

\bibitem{Douglas}
 J.~N.~Douglas,	 F.~N.~Bash,  F.~A.~Bozyan, et al.,
AJ~ {\bf 111}, 1945 (1996).

\bibitem{Ficarra}
A.~Ficarra, G.~Grueff, and G.~Tomassetti,
 A\&AS {\bf 59}, 255 (1985).

\bibitem{Korzhavin}
A.~N.~Korzhavin, Astrofiz. Issl. (Izvestiya SAO)  {\bf 9}, 71
(1977).

\bibitem{Majorova}
E.~K.~Majorova, Astrophys. J. Suppl.~ {\bf 65}, 196 (2010).

\bibitem{Verkh_radio}
O.~V.~Verkhodanov in
 {\it Proc. of the 27 Radio Astron. Conf. Current Radio Astronomy Problems}
 (IAA RAS, St. Petersburg, 1997), 
     p. 322.


\end{thebibliography}
\end{document}